# Joint Inference of User Community and Interest Patterns in Social Interaction Networks

Arif Mohaimin Sadri
e-mail: sadri.buet@gmail.com
Lyles School of Civil Engineering
Purdue University, 550 Stadium Mall Drive
West Lafayette, IN 47907, USA

Samiul Hasan
e-mail: samiul.hasan@ucf.edu
Department of Civil, Environmental, and Construction Engineering
University of Central Florida,12800 Pegasus Drive, Orlando, FL 32816

Satish V. Ukkusuri
Lyles School of Civil Engineering
e-mail: sukkusur@purdue.edu
Purdue University, 550 Stadium Mall Drive
West Lafayette, IN 47907, USA

*__Corresponding author:__ Arif Mohaimin Sadri (Lyles School of Civil Engineering, Purdue University, 550 Stadium Mall Drive, West Lafayette, IN 47907, USA; Email: sadri.buet@gmail.com)

## Abstract

Online social media have become an integral part of our social beings. Analyzing conversations in social media platforms can lead to complex probabilistic models to understand social interaction networks. In this paper, we present a modeling approach for characterizing social interaction networks by jointly inferring user communities and interests based on social media interactions. We present several pattern inference models: i) Interest pattern model (IPM) captures population level interaction topics, ii) User interest pattern model (UIPM) captures user specific interaction topics, and iii) Community interest pattern model (CIPM) captures both community structures and user interests. We test our methods on Twitter data collected from Purdue University community. From our model results, we observe the interaction topics and communities related to two big events within Purdue University community, namely Purdue Day of Giving and Senator Bernie Sanders' visit to Purdue University as part of Indiana Primary Election 2016. Constructing social interaction networks based on user interactions accounts for the similarity of users' interactions on various topics of interest and indicates their community belonging further beyond connectivity. We observed that the degree-distributions of such networks follow *power*-law that is indicative of the existence of fewer nodes in the network with higher levels of interactions, and many other nodes with less interactions. We also discuss the application of such networks as a useful tool to effectively disseminate specific information to the target audience towards planning any large-scale events and demonstrate how to single out specific nodes in a given community by running network algorithms.

## 1. Introduction

The emergence of online social media such as Facebook, Twitter etc. have created ubiquitous social environments. Users can interact with such systems by being friends with others, updating statuses, posting interesting links, mentioning other users in their statuses or posts, commenting or liking others' posts, privately communicating with their connections and in many other ways depending on the type of the system. User interactions within these systems help to construct a network of user relationships representing links of direct social influence. In such a network, two users are connected if they interact with each other or establish a friendship between them. Thus a *social interaction network* is defined as a network of nodes and links, where nodes consist of the users of a particular online social media system and links are established if two nodes have some form of interaction between them.

An individual's connections and activities in a social interaction network enable us to understand the social influence on real world actions. Such knowledge is invaluable for predicting human actions in real world through a social network amplification (Korolov et al., 2015; Kryvasheyeu et al., 2016). User activity in social media has shown its prevalence in recent years, for instance every second over 143K tweets are being generated on Twitter (Krikorian, 2013). As such the world population is more connected and reachable now than ever before and the ease in information sharing and the ability to instigate others have primarily contributed to such connectivity all over the world. However, the purpose and context of the online activity on social media platforms, such as Facebook and Twitter, may vary from user to user. For example, users' check-in activity can be referred as distinct from what users' post or share to disseminate any specific information. One specific feature of such information sharing activity is the ability of users' in mentioning (direct mentions, retweets and replies) others that they follow. Twitter shows both the characteristics of a social network and an informational network (Myers et al., 2014) and users can share short messages up to 140 characters along with the ability to follow other. While the information network properties of Twitter instigate information contagion globally, the social network properties allows access to geographically and personally relevant information (Kryvasheyeu et al., 2016).

Studies have started to acknowledge the potential and need to utilize the rich information on user activities and networks that social media systems provide. The benefits to collect, analyze and use such large-scale and rich information from online information sources have been realized (Lazer et al., 2009). For instance, many researchers used Twitter to study the service characteristics (Guy et al., 2010; Li and Rao, 2010), retweeting activity (Kogan et al., 2015; Starbird and Palen, 2010), situational awareness (Power et al., 2014; Vieweg et al., 2010), virality prediction (Weng et al., 2013) , online communication of emergency responders (Hughes et al., 2014; St Denis et al., 2014), text classification and event detection (Caragea et al., 2011; Earle et al., 2012; Imran et al., 2013; Kumar et al., 2014; Sakaki et al., 2010), devise sensor techniques for early awareness (Kryvasheyeu et al., 2015), human mobility (Wang and Taylor, 2014, 2015), and disaster relief efforts (Gao et al., 2011). Recently, transportation researchers used these data sources extensively for problems related to human mobility patterns (Hasan et al., 2013), origin-destination demand estimation (Cebelak, 2013; Chen and Mahmassani, 2016; Jin et al., 2014; Lee et al., 2016b; Yang et al., 2014), activity-pattern modelling (Hasan and Ukkusuri, 2014, 2015; Lee et al., 2016a; Zhao and Zhang, 2016), social influence in activity patterns, travel survey methods (Abbasi et al., 2015; Maghrebi et al., 2015), transit service characteristics (Collins et al., 2013), and crisis informatics (Ukkusuri et al., 2014) among others. Although a recent study explored user interaction and social influence for tobacco-oriented social networks in social media (Liang et al., 2015), the empirical literature does not provide any specific evidence on how to systematically extract user communities of social interaction networks from social media or similar social sharing platforms..

Understanding social interaction networks has various applications such as effective information dissemination in local communities and managing planned special events. Information dissemination is the proactive distribution of information and spreading awareness to every individual within a community. It requires systematic planning, collection, organization, and delivery techniques before circulating to the target audience using different media and communication means. Successful spreading of awareness to every individual in a community depends on an effective information dissemination process (Cutter and Finch, 2008; Helbing, 2013; Vespignani, 2009). Information dissemination thus constitutes an important and critical factor for the success of organizing Planned Special Events (PSE) that include large sporting events, concerts, conventions and similar events at specific venues such as stadiums, convention centers and others. Because of specific locations and times of occurrence, PSEs are associated with operational needs that can be anticipated and managed in advance(Latoski et al., 2003).

This paper *presents methods for inferring user communities and interests to characterize social interaction networks from online social media data*. By analyzing human interactions in social media platforms, it is now possible to develop complex probabilistic models to identify communities and topics of human social interactions. These communities of direct social influence would be important channels for a number of key stakeholders to target and effectively disseminate information for managing events in real-time. To the best of our knowledge, this is the first study to propose a model that can construct the social interaction network by predicting communities of direct influence. This data-driven approach accounts for the similarity of users' activity and community information for any given event and uses online social media data. For instance, this network of direct social interactions emergency officials to efficiently disseminate specific information to the target audience and better conduct PSEs or similar large-scale events. The contributions of this study are listed below:

- We present the concept of social interaction networks using large-scale social media data
- We present three models to characterize social interaction networks:
    a) *Interest Pattern Model* (IPM) infers well defined patterns of user interests based on their interactions at an aggregate level.
    b) *User Interest Pattern Model* (UIPM) captures user-specific variations in the interest patterns. However, UIPM is only limited to inferring user-specific patterns and does not account for the user connectivity.
    c) *Community Interest Pattern Model* (CIPM) jointly infers user communities and their interaction patterns by taking both user connectivity and topics of interest into account.
- We demonstrate how to target specific nodes within the identified communities by running network models proposed in the network science literature.

## 2. Data Description and Analysis

In this study, we used Twitter REST API searching for the keyword 'purdue' and obtained 56,159 tweets over a month. Each tweet consisted of several words and user mentions which co-appeared with the keyword 'purdue' and resembled higher likelihood of a Twitter subscriber belonging to the Purdue University community. During the course of data collection, we observed two big events, namely *Purdue Day of Giving* and *Senator Bernie Sanders's* visit to Purdue University as part of Indiana Primary Election 2016. Next, we construct the 'purdue' specific online social interaction networks by considering the user mentions. In Twitter, users can post tweets up to 140 characters and each data point can be stored as a tuple Tweet once collected with the following information:

*Tweet (tweet_id) = {tweet, tweet_created_at, user_id, user_screen_name, user_location, user_name, user_followers_count, user_friends_count, user_statuses_count, user_favourites_count,, user_listed_count, user_mention, tweet_retweeted, tweet_lat, tweet_lon}*

For this study, we are interested in using a sub-tuple tweet to infer the links of direct influence that finally evolves into a highly connected network of a given context:

*tweet (tweet_id) = { user_id, tweet, user_mention}*

Let us consider the following three tuples from the tweets generated on Twitter on 04/28/2016 (02:45:40 +0000), 05/02/2016 (14:45:33 +0000) and 05/03/2016 (13:50:21 +0000), respectively.

***tweet(i)*** *= {**709920419529281537**, 'at purdue university, we are in this campaign to win and become the democratic nominee. - bernie sanders htt…', [null]}*
***tweet(ii)*** *= {**3239853627**, 'rt @saracohennyc at purdue university, we are in this campaign to win and become the democratic nominee. - bernie sanders htt…', [**709920419529281537**]}*
***tweet(iii)*** *= {**325069363**, 'rt @bernielovesall: rt @saracohennyc at purdue university, we are in this campaign to win and become the democratic nominee. - bernie sanders htt…', [**3239853627**, **709920419529281537**]}*

Based on the above tweets, there is a directed link from user *709920419529281537* to user *3239853627* and from user *709920419529281537* to *325069363*. Please refer to Fig. 1(a) and Fig. 1(b) for the details of this network construction. By running the twitter REST API between April 16, 2016 and May 16, 2016 for four consecutive weeks, we were able to obtain 56,159 tweets for the query 'purdue' after initial data cleaning for non-English tweets and common stop words. 20,645 of these tweets did not include any user mentions, however, the rest of the tweets included at least one user mention in each tweet (see Fig. 2 to see the number of tweets collected for each day). The differences in the amount of user mentions in the tweets over days are also presented in Fig. 2. It can be clearly seen that the number of tweets having user mentions is almost twice as the number of tweets without mentions every day. These tweets primarily contribute to the formation of networks of direct social influence i.e. social interaction networks. The top 100 most frequent words contribute approximately 32.5% of all the words that appeared in the collected data. Analyzing the dataset, we have identified 34,363 unique users and 38,442 unique undirected links (39,709 links if direction is considered). The undirected network data (observed) is presented in Fig. 4. The degree distributions of such network follows '*power law'* which is indicative of the existence of fewer nodes having higher levels of interactions and many other nodes having very few interactions. This represents the scale-free property of many real networks (Fig. 1 (c)-(d)). We also determine the existence of common users and links (mentions) in each day of data collection with respect to the most active day (May 3, 2016). From Fig. S1 we observe that a minimum of 155 and a maximum of 462 users were active on Twitter for a given day out of 4,192 unique users that were active on May 3. Likewise, Fig. S2 suggests, a minimum of 17 and a maximum of 117 common links were observed in the four consecutive weeks with respect to the most active day when a total of 3,750 user mentions have been observed. We present the heatmap over time using logarithmic color scales for the top 50 most frequent words in Fig. 3. The annual big event *Purdue Day of Giving* was on April 27, 2016 and *Senator Bernie Sanders'* visited Purdue University on the same day. Relevant keywords such as 'purduedayofgiving' and 'sanders' showed highest intensity on April 27, 2016 (Fig. 3).

The dataset selection was due to its relevance in the study of information sharing patterns of specific topics related to Purdue University. Special events affect this behavior over time. Different combinations of words appear in specific topics on which users interact with one another using the user mention feature of Twitter. The frequently appeared words, as presented in Fig. 3, also suggest the emergence of event specific topics such as *Purdue Day of Giving* and *Senator Bernie Sanders*'s visit during *Indiana Primary* among others. Celebrity players of Purdue such as *Anthony Brown* (football), *Spike Albrecht* (basketball) and others also contributed to the development of many topics. For details, please see our discussion on IPM results.

## 3. Modeling Approach: Formulation and Estimation

In this paper, we present several models for inferring user interests and communities from the interactions found in social media data: i) Interest pattern model (IPM) infers user interest patterns at an aggregate-level, ii) User interest pattern model (UIPM) accounts for the user-specific variations in the interest patterns only based on words or text, and iii) Community interest pattern model (CIPM) can jointly infer user communities and interest patterns based on both words and users mentioned in the data. Our hypothesis is that interactions in social media platforms can be described through representative topics which are the distributions of specific words. To infer these topics, we use topic models or Latent Dirichlet Allocation (LDA) (Blei et al., 2003) which have been widely used in machine learning research. While dealing with unlabeled text data that requires natural language processing, LDA is a highly acceptable framework in the machine learning literature as compared to other predictive or classification algorithms such as Support

Vector Machine, Naive Bayes among others. The main idea here is to find the emergence of different topics in a large corpus of documents. Thus, each document can be modeled as a mixture of topics and each topic as a distribution of words (words being the smallest data unit). We present our CIPM formulation as a significant extension to LDA to be able to predict social interaction communities and user interest pattern simultaneously. We also present a Gibbs sampling approach to estimate the model parameters.

A topic model is a generative model that follows a probabilistic process for generating documents based on a set of straightforward probabilistic sampling rules. This process explains how words in a document can be generated based on some latent variables i.e. topics that evolve in a document. The overall procedure includes the following steps (Steyvers and Griffiths, 2007): (i) select a distribution over topics to make a new document, (ii) for each word in the document, select a topic randomly and then a word from that topic following that distribution. The process can also be reversed and the set of topics that generate the collection of documents can be obtained. The model fitness of such generative models should find the best set of latent variables (i.e. topics of the documents) based on which the observed data can be reasonably explained (i.e. words in the documents). The generative process, as discussed above, does not assume any specific ordering of the words ('bag-of-words' assumption in natural language processing) in a document (Steyvers and Griffiths, 2007) and the frequency of word appearance in a document is the only information that is relevant. The ordering of the words can be useful at times, however, this is not captured by topic models. A topic model can also be applied to other types of discrete data.

### *3.1 Interest Pattern Model (IPM)*

Formally speaking, the problem of identifying various social interaction topics is to determine $K$ latent patterns through $\phi_k$ for $k \in \{1,2,\ldots K\}$ and each topic $\phi_k$ is a distribution of different words. A word is defined as the basic unit of data to be selected from a set of possible words of size $W$, a user tweets about $N$ topics, and the user can contribute to the collection of $M$ tweets. The generative process is summarized below (see Fig. 5a):

1. For each topic, $k \in \{1,2,\ldots K\}$, a distribution over words is selected
$$\phi^{(k)} \sim \text{Dirichlet}\ (\beta)$$
2. For each tweet, $m \in \{1,2,\ldots M\}$,
   a) A distribution over topics is selected
   $$\theta^{(m)} \sim \text{Dirichlet}\ (\alpha)$$
   b) For each word $i$ in tweet $m$
      i. Select a topic $z_i \sim \text{Multinomial}\ (\theta^{(m)}); z_i \in \{1,2,\ldots K\}$
      ii. From topic $z_i$, a word is selected
      $$w_i \sim \text{Multinomial}\ (\phi^{(z_i)}); w_i \in \{1,2,\ldots W\}$$

Now, given $M$ tweets, $K$ topics over $W$ unique words, the main objectives of the inference of topic pattern classifications are:

i. Find the probability of a word $w$ given each topic $k$, $P(w|z=k) = \phi_k^w$ where $P(w|z=k)$ is represented with $K$ multinomial distributions $\phi$ over words of size $W$

ii. Find the probability of a topic $k$ for a word in tweet $m$, $P(z=k|m) = \theta_m^k$ where $P(z|m)$ is represented with $M$ multinomial distributions $\theta$ over $K$ topics.

The above model views the topic pattern as a probability distribution over words and tweeting activities as a mixture of these patterns. From $K$ topics, the probability of **i**-th word in a given tweet $m$ is:

$$P(w_i|m) = \sum_{j=1}^{K} P(w_i|z_i = j)\, P(z_i = j|m) \tag{1}$$

Here, $z_i$ is the latent variable referring to the pattern from which the **i**-th word is drawn, $P(w_i|z_i = j)$ indicates the probability of word $w_i$ under the **j**-th pattern and $P(z_i = j|m)$ is the probability of selecting a word from pattern **j** in the tweet $m$. Intuitively, $P(w|z)$ determines the importance of a word in forming a pattern and $P(z|m)$ determines the prevalence of the pattern in different tweets. The complete model of pattern generation by IPM follows:

$$w_i|z_i, \phi^{(z_i)} \sim \text{Multinomial}\left(\phi^{(z_i)}\right)$$

$$\phi \sim \text{Dirichlet}\,(\beta)$$

$$z_i|\theta^{(m)} \sim \text{Multinomial}\left(\theta^{(m)}\right)$$

$$\theta \sim \text{Dirichlet}\,(\alpha)$$

Here, $\alpha$ and $\beta$ are hyper-parameters for the prior distributions of $\theta$ and $\phi$ respectively. We assume Dirichlet prior distributions which are conjugate to the multinomial distributions.

The joint distribution of words and patterns $P(\boldsymbol{w}, \boldsymbol{z})$ written as:

$$P(\boldsymbol{w}, \boldsymbol{z}) = P(\boldsymbol{w}|\boldsymbol{z})\, P(\boldsymbol{z}) \tag{2}$$

The first term can be written as (Griffiths and Steyvers, 2004):

$$P(\boldsymbol{w}|\boldsymbol{z}) = \left(\frac{\Gamma(W\beta)}{\Gamma\,(\beta)^W}\right)^K \prod_{k=1}^{K} \frac{\prod_w \Gamma(n_k^w + \beta)}{\Gamma(n_k^{(\cdot)} + W\beta)} \tag{3}$$

Here $n_k^w$ is the number of times word $w$ is assigned to pattern $k$ and $n_k^{(\cdot)} = \sum_{w-1}^{W} n_k^w$. The second term can be written as:

$$P(\boldsymbol{z}) = \left(\frac{\Gamma(K\alpha)}{\Gamma\,(\alpha)^K}\right)^M \prod_{m=1}^{M} \frac{\prod_k \Gamma(n_m^k + \alpha)}{\Gamma(n_m^{(\cdot)} + K\alpha)} \tag{4}$$

Here $n_m^k$ is the number of times a word from tweet $m$ is assigned to pattern $k$ and $n_m^{(\cdot)} = \sum_{k-1}^{K} n_m^k$. A pattern can be assigned to a word using the following conditional distribution (Griffiths and Steyvers, 2004) :

$$P(z_i = k|\boldsymbol{z_{-i}}, \boldsymbol{w}) \propto \frac{n_{-i,k}^{w_i}+\beta}{n_{-i,k}^{(\cdot)}+W\beta} \; \frac{n_{-i,m_i}^{k}+\alpha}{n_{-i,m_i}^{(\cdot)}+K\alpha} \tag{5}$$

Here, $n_{-i}$ is the count excluding the current pattern assignment of $z_i$. The first ratio expresses the probability of word $w_i$ in pattern $k$ and the second ratio expresses the probability of pattern $k$ in the tweet $m$.

### *3.2 User Interest Pattern Model (UIPM)*

IPM, as discussed above, when applied over all the users assumes that each tweet consists of topic discussion of a unique user. It does not capture the user-level variation of the evolving topics i.e. the user-level topic patterns. We extend IPM to UIPM to infer the user-specific patterns and the probabilistic generative process of UIPM is summarized below (see Fig. 5b):

1. For each topic, $k \in \{1,2, \dots K\}$, select a distribution over words
   $$\phi^{(k)} \sim \text{Dirichlet}\ (\beta)$$
2. For each user, $u \in \{1,2, \dots U\}$, select a distribution over topics
   $$\theta^{(u)} \sim \text{Dirichlet}\ (\alpha)$$
   a) For each tweet, $m \in \{1,2, \dots M_u\}$ of user $u$
      i. For each word $i$ in the tweet $m$
         A. Select a topic $z_i \sim$ Multinomial $\left(\theta^{(u)}\right)$; $z_i \in \{1,2, \dots K\}$
         B. From topic $z_i$, select a word
            $w_i \sim$ Multinomial $\left(\phi^{(z_i)}\right)$; $w_i \in \{1,2, \dots W\}$

The difference of this model formulation as compared to the previous one is that the second term of Eq. 1 can be rewritten as follows:

$$P(\boldsymbol{z}) = \left(\frac{\Gamma(K\alpha)}{\Gamma\ (\alpha)^K}\right)^U \prod_{u=1}^{U} \frac{\prod_k \Gamma(n_u^k+\alpha)}{\Gamma(n_u^{(\cdot)}+K\alpha)} \tag{6}$$

Here $n_u^k$ is the number of times a word from user $u$ is assigned to pattern $k$ and $n_u^{(\cdot)} = \sum_{k-1}^{K} n_u^k$. A user-specific pattern can be assigned using the following conditional distribution:

$$P(z_i = k|\boldsymbol{z_{-i}}, \boldsymbol{w}, \boldsymbol{u}) \propto \frac{n_{-i,k}^{w_i}+\beta}{n_{-i,k}^{(\cdot)}+W\beta} \; \frac{n_{-i,u_i}^{k}+\alpha}{n_{-i,u_i}^{(\cdot)}+K\alpha} \tag{7}$$

Here, $n_{-i}$ is the count excluding the current pattern assignment of $z_i$. The first ratio expresses the probability of word $w_i$ in pattern $k$ and the second ratio expresses the probability of pattern $k$ in the tweeting activities of user $u$.

### *3.3 Community Interest Pattern Model (CIPM)*

UIPM is limited to only capturing user-specific variations in the emerging topics. It does not capture the community-level variation of the evolving topics i.e. the community-level topic patterns. In this section, we extend the topic model to CIPM where community-specific patterns are represented. CIPM accounts for both community structures and user interests based on both words and users mentioned in the tweets. Each user mentioned in a given tweet by another user contributes to the higher likelihood of these two users belonging to the same community.

The probabilistic generative process for the model is summarized below (see Fig. 5c):

1. For each topic, $k \in \{1,2, \dots K\}$, select a distribution over words
$$\phi^{(k)} \sim \text{Dirichlet}\ (\beta)$$
2. For each community, $c \in \{1,2, \dots C\}$, select a distribution over users
$$\mu^{(c)} \sim \text{Dirichlet}\ (\gamma)$$
3. For each user, $u \in \{1,2, \dots U\}$, select a distribution over topics
$$\theta^{(u)} \sim \text{Dirichlet}\ (\alpha)$$
   a) For each tweet, $m \in \{1,2, \dots M_u\}$ of user $u$
      i. Select a community $c$ for the topic $m$
      ii. For each word $i$ in the tweet $m$
         A. Select a user $x$ from $c$ to be mentioned as observed in the tweet
         B. Select a topic $z_i \sim \text{Multinomial}\left(\theta^{(u)}\right)$; $z_i \in \{1,2, \dots K\}$
            Assign the same topic to $x$
         C. From topic $z_i$, select a word
$$w_i \sim \text{Multinomial}\left(\phi^{(z_i)}\right);\ w_i \in \{1,2, \dots W\}$$

A community-specific pattern can be assigned using the following conditional distribution:

$$P(z_i = k|\boldsymbol{z_{-i}}, \boldsymbol{w}, \boldsymbol{u}, \boldsymbol{c}) \propto \frac{n_{-i,k}^{w_i}+\beta}{n_{-i,k}^{(\cdot)}+W\beta}\ \frac{n_{-i,u_i}^{k}+\alpha}{n_{-i,u_i}^{(\cdot)}+K\alpha}\ \frac{n_{-i,u_i}^{c}+\gamma}{n_{-i,u_i}^{(\cdot)}+C\,\gamma} \tag{8}$$

Here, $n_{-i}$ is the count excluding the current pattern assignment of $z_i$. The first ratio expresses the probability of word $w_i$ in pattern $k$, the second ratio expresses the probability of pattern $k$ in the tweeting activities of user $u$ and the third ratio expresses the probability of user $x$ in the tweeting activities of user $u$.

### *3.4 Parameter estimation*

There are various approximation techniques for estimating the parameters of this model (Blei et al., 2003; Griffiths and Steyvers, 2004). We used the Gibbs sampling approach proposed by (Griffiths and Steyvers, 2004). The algorithm can be found in detail in (Griffiths and Steyvers, 2004). Only a brief description of the approach is provided here. To estimate the model parameters, a Markov Chain Monte Carlo (MCMC) procedure is used. In MCMC, samples are taken from a Markov chain constructed to converge to a target distribution. In our model, each state of the chain is the assignment of a pattern to a word and the transition from one state to another follows a specific rule based on Gibbs sampling approach (Robert, 2007). In this procedure, the next state is reached by sampling the variables from a conditional distribution which specifies the distribution of the variables conditioned on the current assignment of all other variables and the observations. The parameters of IPM, representing the hidden patterns, can be computed as:

$$\hat{\phi}_k^w = \frac{n_k^w+\beta}{n_k^{(\cdot)}+W\beta}\,;\ \hat{\theta}_m^k = \frac{n_m^k+\alpha}{n_m^{(\cdot)}+K\alpha} \tag{9}$$

The parameters of UIPM can be computed as:

$$\hat{\phi}_k^w = \frac{n_k^w+\beta}{n_k^{(\cdot)}+W\beta}\,;\ \hat{\theta}_u^k = \frac{n_u^k+\alpha}{n_u^{(\cdot)}+K\alpha} \tag{10}$$

Finally, CIPM model parameters are computed as:

$$\hat{\phi}_k^w = \frac{n_k^w+\beta}{n_k^{(\cdot)}+W\beta}\,;\hat{\theta}_u^k = \frac{n_u^k+\alpha}{n_u^{(\cdot)}+K\alpha}\,;\ \hat{\mu}_u^c = \frac{n_u^c+\gamma}{n_u^{(\cdot)}+C\,\gamma} \tag{11}$$

We implemented our model formulations by using Python programming language to process the data and estimate model parameters. Cython, an extension of Python, was used to write the core computational steps of Gibbs sampling procedure that reduced the computational time significantly. The number of input words and user mention sequence, the number of latent patterns, and the number of samples determine the actual time required to estimate the model parameters. A typical setup for the input data used in the paper took less than an hour to estimate the IPM and UIPM parameters, however, more than four hours in case of CIPM.

## 4. Results

The model selection was based on *Perplexity* – a metric to measure the predictive capacity of the model to infer the unseen data in each run. In machine learning, *perplexity* is a commonly used metric to report the performance of a probabilistic model that refers to the average likelihood of obtaining a test data set given a set of model parameters. *Perplexity* can be defined as the exponential of the negative of average predictive likelihood of a test data given a model (Griffiths and Steyvers, 2004). In this study, the algorithm was run for different number of topic patterns (K) and perplexity in each run was computed. Next, the optimal number of patterns was selected based on perplexity values. For a given set of words $\{\boldsymbol{w_m}\}$ and $m \in D^{test}$ given a model $\mathcal{M}$, *Perplexity* of a test data set can be defined as:

$$Perplexity = \exp\left[-\frac{\sum_{m=1}^{M}\log p(\boldsymbol{w_m}|\mathcal{M})}{\sum N_m}\right] \tag{12}$$

where $N_m$ is the number of words in each topic $m$ and $p\ (\boldsymbol{w_m}|\mathcal{M})$ can be derived from Eq. (1).

In order to estimate perplexity values, the data set was randomly split with 90% of the users in the training set with the rest in the test set; the parameters of the topic model on the training data set were estimated; and finally the perplexity values on the test data set were computed. While a lower value of perplexity refers to a better model performance, an increase in the number of topics reduces the perplexity and there is no significant improvement beyond a certain number of topics (Fig. S3).

### *4.1 IPM patterns*

Based on perplexity values, K = 100 was selected for running the Gibbs algorithm for finding the topic patterns. Table S1 presents the results of the topic model applied to the data. Although 100 latent patterns were estimated, only a few insightful and interesting patterns are reported here and explained along with the probabilities of the top 10 words in each topic (Table S1):

- Patterns 1, 41, and 48 resemble academic interests within Purdue University community
- Pattern 97 shows relevance to Purdue alumni and their activities
- Pattern 7 reflects user (students) interest related to convocation and graduation which typically occur during mid-May for the Spring semester
- Pattern 11 conveys user (students) concern about tuition fees
- Patterns 43, 59, and 98 express user interest related to recent research accomplishments on campus
- Pattern 23 is the evidence of user gratitude to university donors
- Pattern 2 is specific to the interests about the Online Writing Lab (OWL) at Purdue
- Patterns 5, 71, and 76 capture user interest about the large annual event on campus, namely *Purdue Day of Giving*
- User interactions related to *Senator Bernie Sanders'* visit during *Indiana Election Primary 2016* are captured by Patterns 78, 82, and 85
- Patterns 12, 15, 17, 33, 36, and 64 resemble campus-wide online discussions about celebrity players or influential people who are affiliated with *Purdue University*
- Patterns 28, 35, 53, 81, 14, and 77 are specific to weekend games on campus. These include team mascots, upcoming matches, opponent teams, joining of new players, big ten conference, among others
- Patterns 19, 24, 34, 66, 86 and 90 are some well-defined, however, irrelevant topics related to *Purdue Pharma* which is a privately held pharmaceutical company located in *Stamford, Connecticut*.

IPM results indicate that the proposed model is a useful tool to find the latent patterns of user interests and interactions in social media. The observed patterns primarily contain the campus-wide activities with a higher concentration of user interactions about regular weekend games. The model can capture unique special events or annual big events. IPM thus also allows to eliminate irrelevant social interactions. Traditional survey techniques hardly can infer such user interest metric based on social interactions.

### *4.2 UIPM patterns*

In this section, we present the results from user interest pattern model (UIPM). Similar to IPM model selection, we use perplexity values to determine the number of interest patterns to be estimated. Fig. S3 presents the results for perplexity measurements for UIPM. Based on these perplexity values we select K = 100 for estimating the model parameters. Table S2 presents a few of the user interest patterns estimated and the probability of the top 10 words for each pattern reported. We discuss about patterns that are specific to the two large and unique events that we observed during the course of data collections (Fig. 6):

- Patterns 1, 21, and 56 resemble user interactions about *Senator Bernie Sanders'* visit during *Indiana Election Primary 2016*
- While it appears users like *41621505, 16440677, 2228806220* and others primarily talk about the outcome of the event (Pattern 1) i.e. *Senator Sanders'* winning the primary election, users such as 16664309, 721428816971894000, 123269504 and others interacted were likely to interact long before the event (Pattern 56). On the other hand, Pattern 21 is indicative of user interactions the event.
- Patterns 19, 20, and 25 capture user interaction about the annual large campus event *Purdue*

*Day of Giving*

- While Pattern 19 demonstrates user interactions before the event i.e. supporting the cause of the event, pattern 20 suggests user interactions after the event since the users discussed abut record breaking grant collected this year.

Similar to IPM, as discussed in the previous section, these user-level patterns demonstrate the applicability of our modelling approach to extract the hidden underlying patterns of user-specific interests. The reported patterns contain the likely interests of the users on various social interactions. We also report the top users for each of the interest patterns along with their corresponding probabilities. These users indicate the top contributors for the corresponding patterns. It is also important to note here that a user's social interactions can be modelled as a mixture of patterns since the user can have varied interests that belong to few specific patterns and a user can iterate the same pattern several times. In addition to the word proportions for a given pattern, the UIPM can also determine user-specific pattern proportions ($\hat{\theta}_u^k$). These pattern proportions can be used to find the similarity among the users based on their interest and social interactions. From the subgraph visualizations of the top 10 users for each pattern listed in Table S2, we observe in Fig. 6 that these users are highly connected to each having more weighted social interactions. This justifies the exclusive purpose of considering the user mentions i.e. links of social interaction networks which is why we extend our UIPM formulation to CIPM that can obtain community specific pattern as discussed in details in the next section.

### *4.3 CIPM patterns*

In this section, we present the results from community interest pattern model (CIPM). Similar to IPM and UIPM model selection based on perplexity values (Fig. S3), we select K = 100 for estimating the CIPM model parameters and assign the users into 40 different communities to explain the community patterns (18,077 users, K=100, C =40, p>=0.025). For 100 topics, we run the CIPM for different number of communities and present the variation in the number of users that can be assigned into C different number of communities with probability greater or equal to 1/C (Fig. S4). We observe that maximum number of users are assigned when K = C = 100. Although a user can belong to any community with certain probability, however, we set the minimum probability threshold as 1/C in order to assign users in various communities. In order to better analyze the community patterns inferred by CIPM, we visualize and discuss about the largest connected components (LCC) of two community patterns (community 2 and community 3) for the user assignment to 40 communities in which the maximum probability that a user can belong is considered (Please see Fig. 7(a) for the distribution of user assignment in communities).

Table S3 presents the community interest patterns estimated and the probability of the top 10 words for each interest pattern to which users, assigned to community 2 (C-2) and community 3 (C-3), had their maximum contribution. The LCC's of the top 4 communities: C-1, C-21, C-3 and C-2 (according to the size, respectively) are presented in Fig. 7(b). In order to present CIPM results, we discuss below some key community and interest patterns for the top 4 communities:

- All the users, assigned to any given community, belong to their respective communities with at least 2.5% probability. These assignments are based both on topic similarity and mention similarity
- C-1 LCC and C-21 LCC respectively include 4.21% and 3.96 % of all the users (21,045) that exist in the LCC of the original graph (please refer to the center of Fig. 4)
- Almost all the users assigned in C-2 LCC primarily contributed to Pattern 1 (captures overall interest about *Purdue University* and its logo *BoilerUp*) except for user *714816804858630144* who had interest more towards Pattern 5. However, all these users belong to the same community (C-2) because of their mention similarity i.e. high connectivity as presented in Fig. 7(b)

- Likewise, 32 users belong to belong to C-3 LLC where many users also expressed primary interest about Pattern 1 and some contributed to other topics such as Pattern 23, 62, 91 and 99)

As an extension to UIPM, as discussed in the previous section, our CIPM formulation is capable of capturing community-level patterns by accounting for the hidden underlying patterns of user-specific interests. The reported patterns contain the likely interests of the users on various social interactions and their likelihood of belonging to certain communities as inferred from their user mentions. In addition to the word proportions for a given pattern, the CIPM can also determine community-specific pattern proportions ($\hat{\mu}_u^c$). These pattern proportions can be used to find the similarity among the users based on their connectivity and social interactions. From the subgraph visualizations of the top 4 communities, we observe in Fig. 8a that several users in C-3 LCC are not only connected by user mentions but also highly weighted. This demonstrates how CIPM can jointly infer user communities and interests in social interaction networks. CIPM can also classify nodes in a large graph based on their community assignment as presented in Fig. S5.

## 5. Applications

The interdependence between complex networks having dynamic, irregular structure and the functional behavior of the network agents has significant outcomes when the robustness and resilience of a real network is considered and the way networks respond to targeted failure due to external disturbances as suggested in the *Network Science* literature (Albert and Barabási, 2002; Albert et al., 2000; Boccaletti et al., 2006; Newman, 2010; Newman, 2003). The prevalence of networked systems has resulted in several studies with applications in various domains over the last decade. A few examples of such studies may include travel demand network (Saberi et al., 2016), the Commonwealth trade network (Ukkusuri et al., 2016), contagion of risk perception during crisis (Hasan and Ukkusuri, 2011), disease transmission (Anderson et al., 1992; Murray, 2002), email networks and computer virus transmission (Balthrop et al., 2004; Newman et al., 2002), power grid networks (Kinney et al., 2005; Sachtjen et al., 2000), market disruptions (Sornette, 2009), information propagation (Coleman et al., 1966), and many others. Such studies primarily focused on exploring the emergence of new innovations or ideas based on agent interactions, identifying influential players in the network, maximizing network influence based on certain mechanism, determining under what conditions contagions become global cascade among others.

In this section, we demonstrate the benefits of inferring communities in social interaction networks, being more specific to the transportation research domain. Effective information dissemination is a key to successfully arrange *Planned Special Events (PSE)*, briefly discussed as our motivation in Section 1. Organizing PSEs have several challenges including parking management, crowd management, pedestrian facility design, and special facility for senior citizens and handicapped individuals, providing transit facility for captive riders among others. In addition, police enforcements often need to close several streets for security reasons, manage crowds who walk together to the location and guide motorists to specific routes who are unfamiliar with the area. Individuals attending these events travel by various travel modes, i.e. walk, private car and public transit. However, despite several operational needs and technical requirements to manage PSEs, the empirical literature does not provide any guidance to local traffic managers and emergency response personnel to identify targeted groups or communities and disseminate travel specific information.

For example, by running relevant network models for Community 3 (see previous section for details), we report both graph-level and node-level properties in Table 1. Node level properties are important to understand the role of different nodes (network agents) on the information propagation at a local scale and identify the nodes that can play important role during a crisis or emergency because of their higher access to many other nodes. The ego node of the largest hub, user *615833064* in C-3 with degree score 22, depicts its influential position in the largest connected component of C-3 (Fig. 8b) when accessibility from a given node to many other nodes is considered. This user account on Twitter belongs to *Spike Albrecht* who is a college basketball player from Crown Point, Indiana and he will play for the Purdue Boilermaker Team for the 2016-17 season. While a node degree is the number of edges adjacent to that node, average neighbor degree refers the mean degree of the neighborhood. On the other hand, transitivity implies that two nodes are highly likely to be connected in a network, given each of the nodes are connected to some other node. This is indicative of heightened number of triangles (sets of three nodes each of which is connected to each of the others) that exist in real networks (expressed in terms of clustering coefficient) (Newman, 2003). In case of social networks, transitivity refers to the fact that the friend of one's friend is likely also to be the friend of that person. This property is dominant for user *1180961940* in C-3 although it has very low degree as compared to user *615833064*.

*Network Density* is a graph-level property that is frequently used in the sociological literature (Scott, 2012). The density is 0 for a graph without any link between nodes and 1 for a completely connected graph. While the eccentricity of a node in a graph is the maximum distance (number of steps or hops) from that node to all other nodes; radius and diameter are the minimum and maximum eccentricity observed among all nodes, respectively. For C-3, we observe that the radius is 3 and diameter equals 5. Centrality measures indicate how central a given node in the network. Betweenness centrality of a node is the sum of the fraction of all-pairs of shortest path that pass through that node (Brandes, 2001, 2008; Brandes and Pich, 2007). Closeness centrality of a node is the reciprocal of the sum of the shortest path distances from node to all other nodes in the graph. Closeness is normalized by the sum of minimum possible distances of all other nodes since the sum of the distances depend on the number of nodes in the graph (Freeman, 1978) for details. Higher values of closeness imply higher centrality and user *615833064* is the most central node in the largest connected component of C-3 (Table 1).

The properties of social interaction networks, as observed in this study, have fundamental implications towards effective information dissemination. For example, removal of hubs (high degree nodes) would cause major disruption and network agents would fail to communicate since the regular length of path will increase because of many disconnected pairs of nodes. For any *Planned Special Event (PSE)*, the assembling of vehicles and pedestrians in a short amount of time cause transportation and transit authorities to often encounter significant challenges in controlling the induced traffic coming from different origins before the event and departing from the event location after the event.

## 6. Conclusions

In this study, we are interested in jointly modeling the interests and social interactions among users in a social sharing platform within a unified modeling framework. The work is motivated by the observation that a link between two users is not only determined by interest similarity, but also affected by the community ties between the users. Indeed, in-town users are more likely to be influenced by their friends (as compared to those out-of-town), and researchers are more likely to cite papers presented at the conferences they attend or in the journals they read. This is evident since users are inherently more aware of the activity in their community and might not be aware of relevant activity outside it. By accounting for both topic similarity and social network relations, we can better identify the reasons for the presence or absence of a link, and, in turn, find improved user interaction topics and communities.

This paper demonstrates the uses of large-scale data available from different social sharing platforms to characterize and measure social network influence. We observe that user interactions in such networks follows *power-law* indicating fewer nodes in with higher levels of interactions and many other nodes with less interactions. We develop several pattern inference models: *i) Interest pattern model (IPM)* infers population-level user interest patterns, *ii) User interest pattern model (UIPM)* accounts for user-level variations of interests based on the texts generated from social interactions, and *iii) Community interest pattern model (CIPM)* jointly infers user communities and interests based on text and users mentioned in the tweets. These models are expected to leverage the process of information dissemination in targeted communities by having both the knowledge user interest and interactions in a social sharing environment. Predicting user communities will allow local emergency planners to implement proactive event management plans across different special events. These networks of direct social influence can further be used to conduct different network experiments such as influence maximization, community detection, identifying influential nodes and many others.

The study is limited to a university-specific data and the patterns are explained for the communities and user interests that belong to only Purdue University. The data also suffers from lack of representativeness and sampling biases due to the limited collection period of one month. We also note here the scalability and computational issues of our proposed algorithms for real-world applications given the large size of the data. However, the methodologies presented in this study would be useful to predict user communities based on their interaction and interests in various social sharing platforms.

## Acknowledgements

The authors are grateful to National Science Foundation for the grant CMMI-1131503 and CMMI-1520338 to support the research presented in this paper. However, the authors are solely responsible for the findings presented in this study.

## Author Contributions Statement

All the authors have contributed to the design of the study, conduct of the research, and writing the manuscript.

## Additional Information

**Competing financial interests:** Authors declare no competing financial interests.

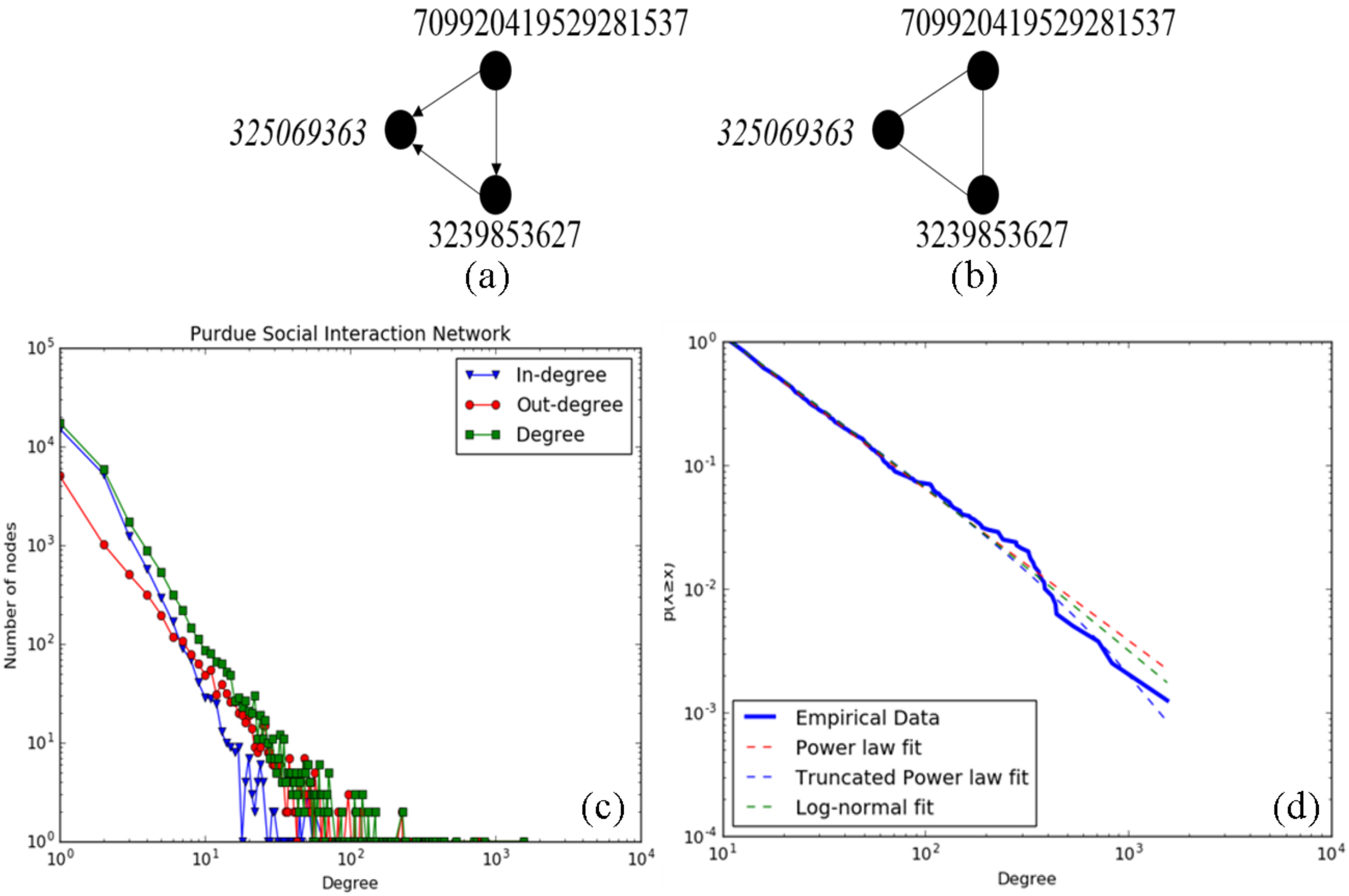

Fig. 1. Construction and Visualization of Social Interaction Network. (a) Directed graph, (b) Undirected graph, (c) In-degree, Out-degree and Degree Distributions, (d) Comparison of data fitting with different distributions

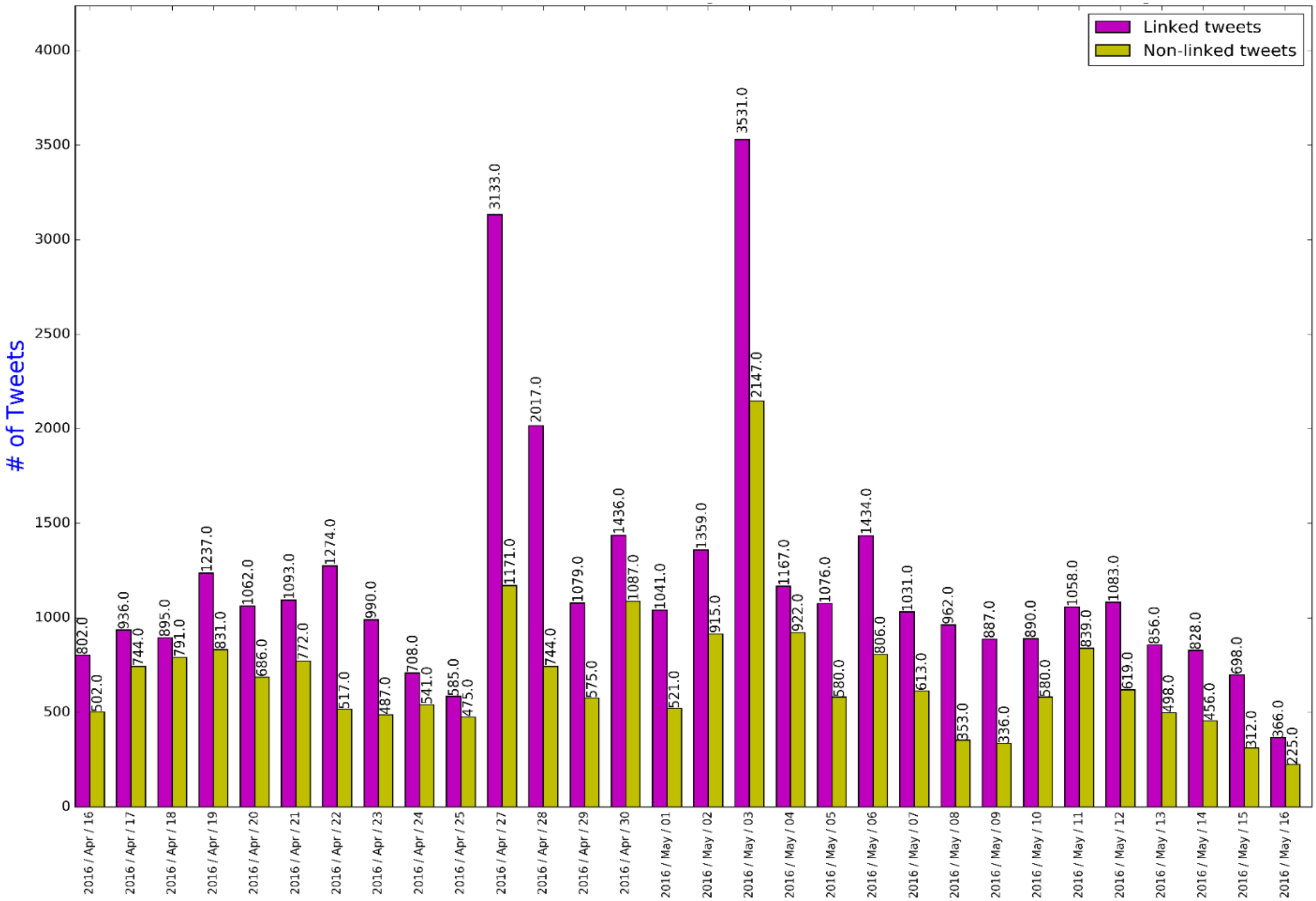

Fig. 2. Number of tweets collected for each day for the data collection period (April 16, 2016-May 16, 2016)

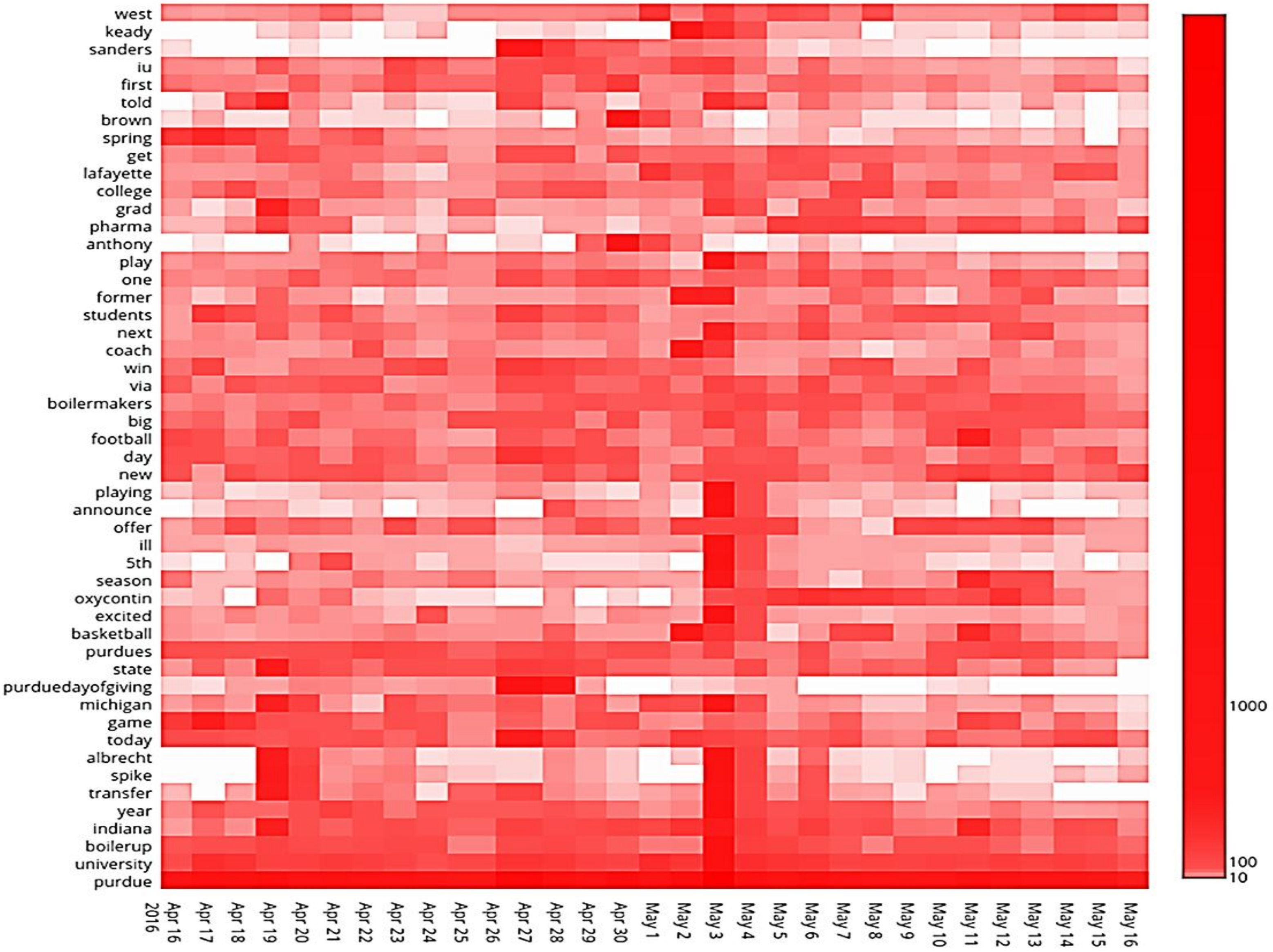

Fig. 3. Word appearance over time (word frequency rank: 1-50)

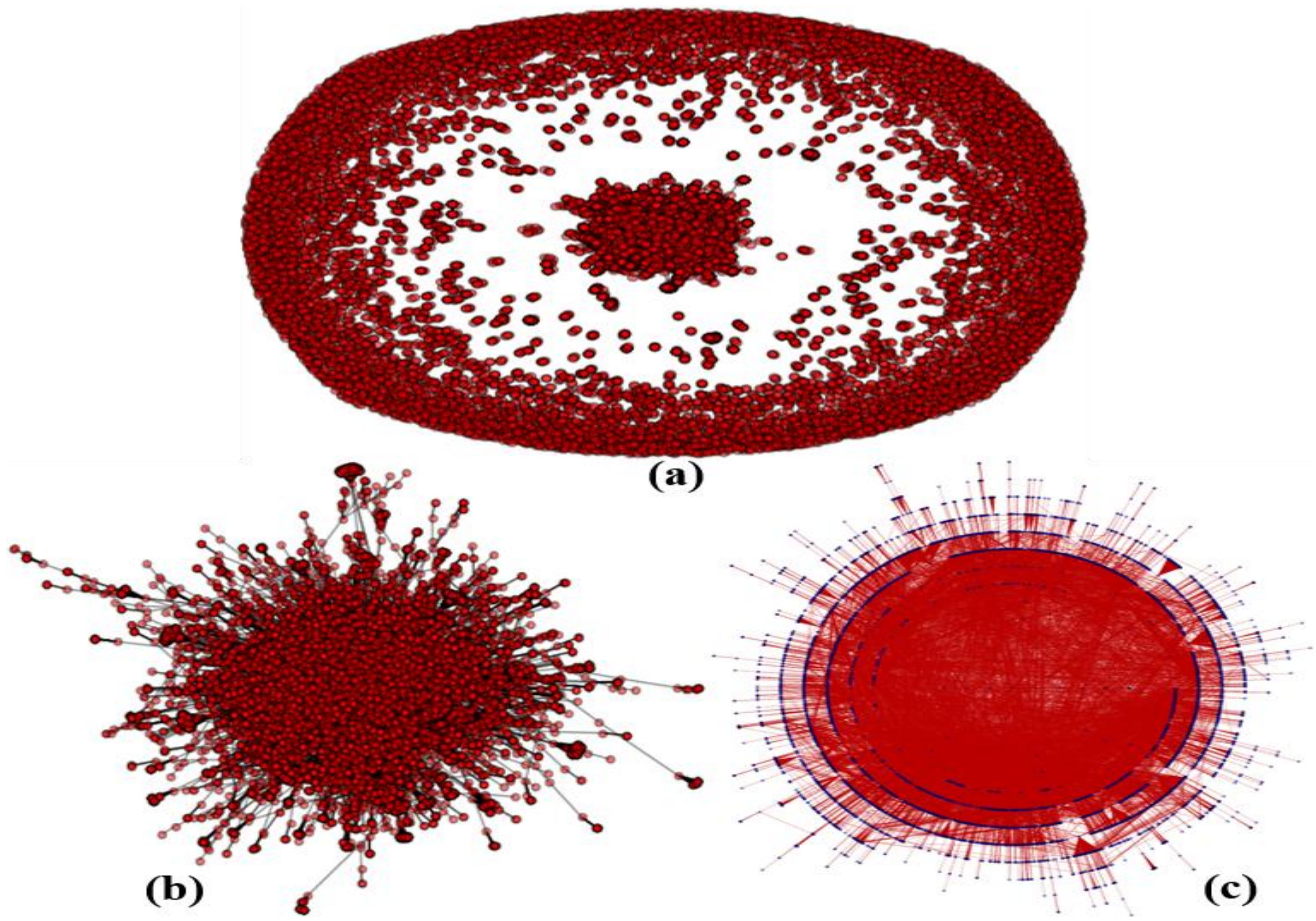

Fig. 4. Social interaction network data based on user mentions (a) undirected graph, (b) spring layout of the largest connected component (lcc), (c) circular tree layout of lcc

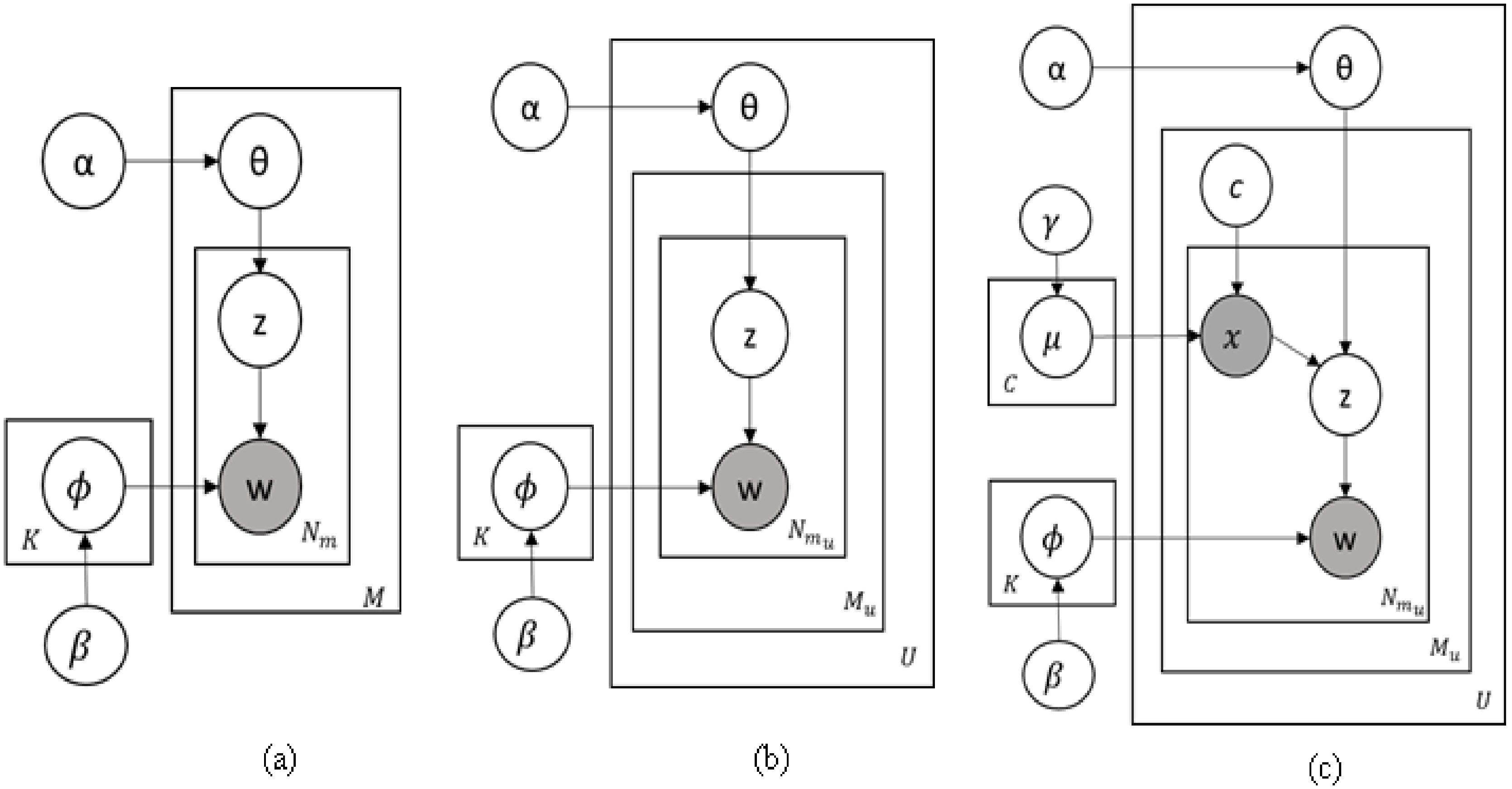

Fig. 5. Graphical representation of probabilistic generative process (a) IPM, (b) UIPM, (c) CIPM. (rectangle plates represent the repetitiveness of the data, white circles present random variables and shaded circles represent observed variables, arrows represent the dependency between different entities.)

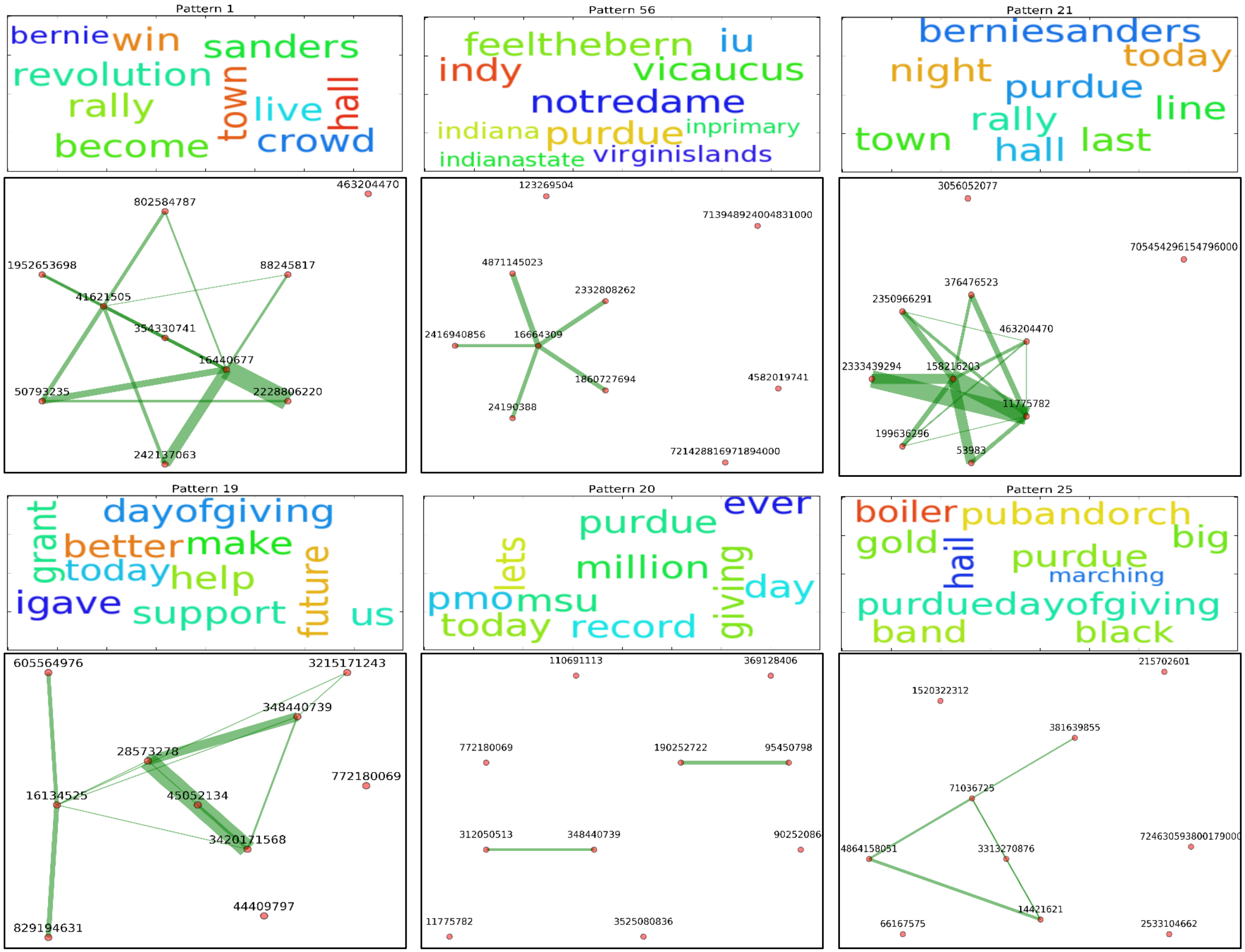

Fig. 6. Higher connectivity of top users in different UIPM patterns

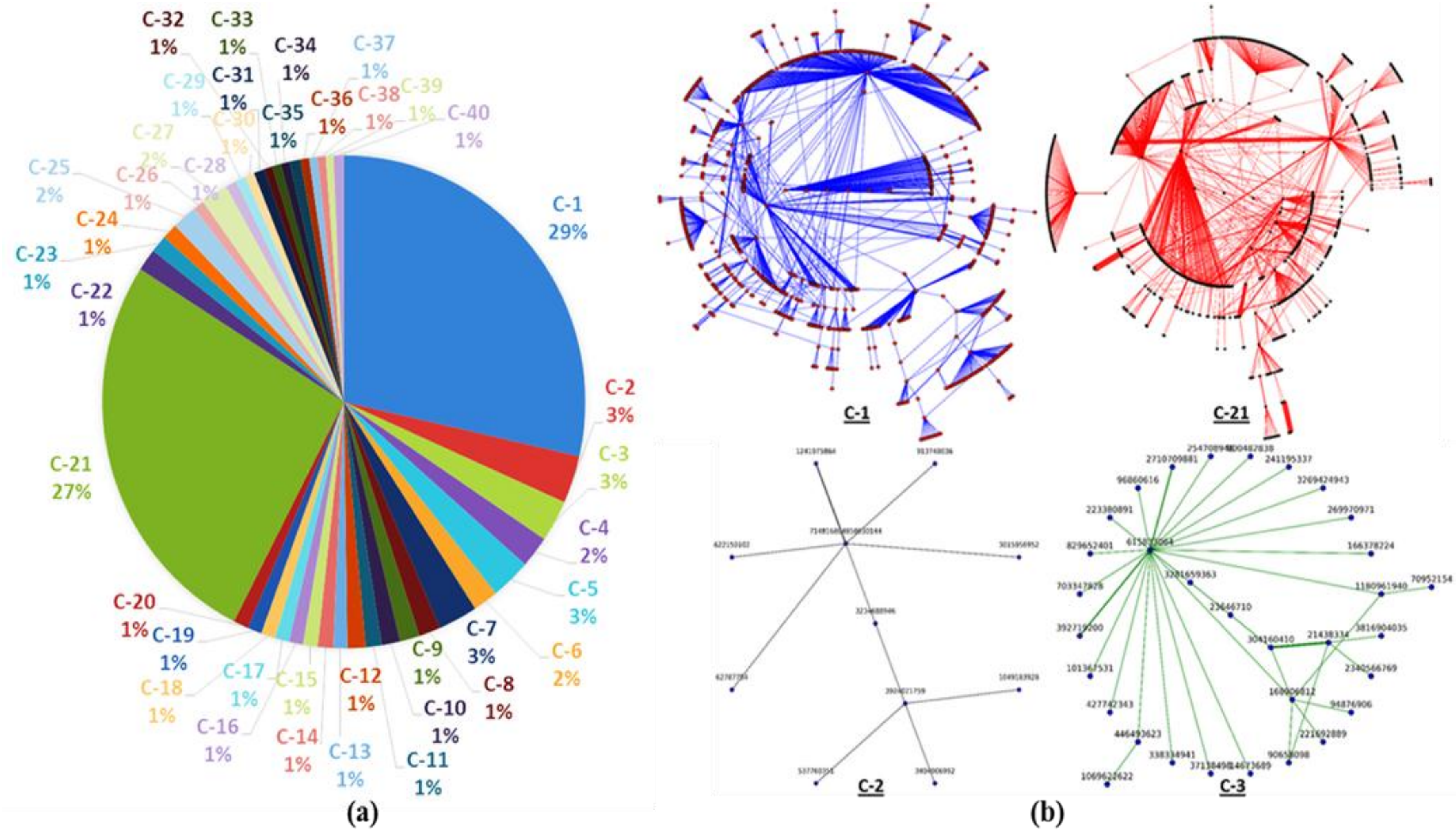

Fig. 7. CIPM Patterns (a) Distribution of users in different communities, and (b) LCC of Top 4 Communities: C-1, C-21, C-2 and C-3.

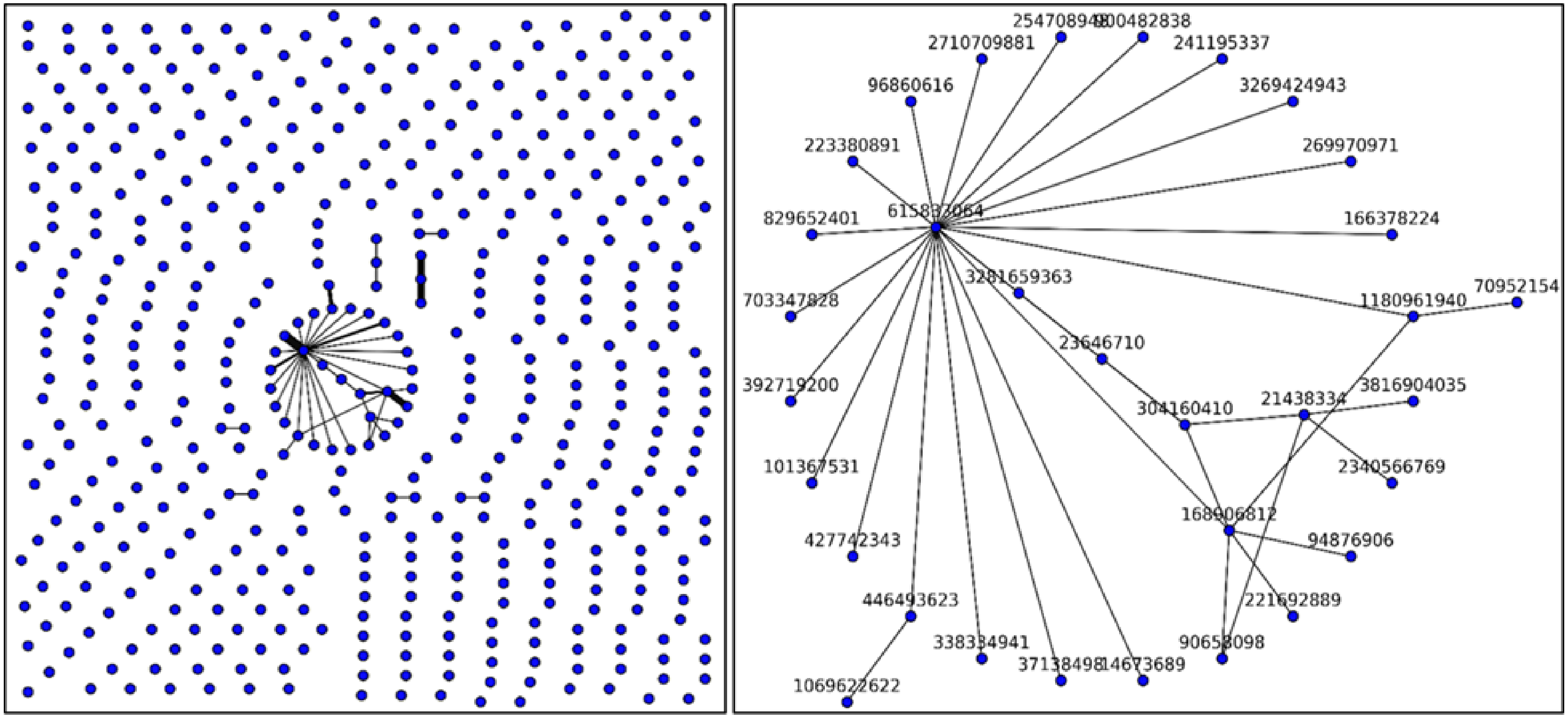

(a) (b)

Fig. 8: Social Interaction Community 3 (C-3). (a) undirected graph (492 nodes, 44 links, 452 connected components, 444 isolates, and density 0.00036), (b) undirected largest connected component, lcc (32 nodes, 34 links, and density 0.00015); user id 615833064 of the largest hub belongs to Spike Albrecht, a transferred basketball player who will be playing for Purdue team for the 2016-17 season.

Table 1: Node-level network properties of community 3 largest connected component

| **User Id** | ***Degree*** | ***Clustering Coefficient*** | ***Eccentricity*** | ***Average Neighbor Degree*** | ***Betweenness Centrality*** | ***Closeness Centrality*** |
|---|---|---|---|---|---|---|
| 304160410 | 3 | 0 | 3 | 9.3 | 0.177 | 0.492 |
| 223380891 | 1 | 0 | 4 | 22 | 0 | 0.431 |
| 829652401 | 1 | 0 | 4 | 22 | 0 | 0.431 |
| 2340566769 | 1 | 0 | 5 | 4 | 0 | 0.272 |
| 94876906 | 1 | 0 | 4 | 5 | 0 | 0.341 |
| 703347828 | 1 | 0 | 4 | 22 | 0 | 0.431 |
| 392719200 | 1 | 0 | 4 | 22 | 0 | 0.431 |
| 23646710 | 2 | 0 | 4 | 2.5 | 0.004 | 0.341 |
| 101367531 | 1 | 0 | 4 | 22 | 0 | 0.431 |
| 427742343 | 1 | 0 | 4 | 22 | 0 | 0.431 |
| 168906812 | 5 | 0.100 | 3 | 5.8 | 0.183 | 0.508 |
| 446493623 | 2 | 0 | 4 | 11.5 | 0.065 | 0.443 |
| 3816904035 | 1 | 0 | 5 | 4 | 0 | 0.272 |
| 70952154 | 1 | 0 | 5 | 3 | 0 | 0.323 |
| 338334941 | 1 | 0 | 4 | 22 | 0 | 0.431 |
| 37138498 | 1 | 0 | 4 | 22 | 0 | 0.431 |
| 21438334 | 4 | 0 | 4 | 1.8 | 0.131 | 0.369 |
| 14673689 | 1 | 0 | 4 | 22 | 0 | 0.431 |
| 615833064 | 22 | 0.004 | 3 | 1.5 | 0.901 | 0.738 |
| 1180961940 | 3 | 0.333 | 4 | 9.3 | 0.065 | 0.470 |
| 166378224 | 1 | 0 | 4 | 22 | 0 | 0.431 |
| 269970971 | 1 | 0 | 4 | 22 | 0 | 0.431 |
| 1069622622 | 1 | 0 | 5 | 2 | 0 | 0.310 |
| 90658098 | 2 | 0 | 4 | 4.5 | 0.026 | 0.373 |
| 221692889 | 1 | 0 | 4 | 5 | 0 | 0.341 |
| 3269424943 | 1 | 0 | 4 | 22 | 0 | 0.431 |
| 3281659363 | 2 | 0 | 4 | 12 | 0.027 | 0.443 |
| 241195337 | 1 | 0 | 4 | 22 | 0 | 0.431 |
| 900482838 | 1 | 0 | 4 | 22 | 0 | 0.431 |
| 254708948 | 1 | 0 | 4 | 22 | 0 | 0.431 |
| 2710709881 | 1 | 0 | 4 | 22 | 0 | 0.431 |
| 96860616 | 1 | 0 | 4 | 22 | 0 | 0.431 |

# Supplementary Information

Joint Inference of User Community and Interest Patterns in Social Interaction Networks

Arif Mohaimin Sadri[1,*], Samiul Hasan[2], Satish V. Ukkusuri[1]

[1] Lyles School of Civil Engineering, Purdue University, 550 Stadium Mall Drive, West Lafayette, IN 47907, USA

[2] Department of Civil, Environmental, and Construction Engineering, University of Central Florida, 12800 Pegasus Drive, Orlando, FL 32816.

[*]**Corresponding author:** Arif Mohaimin Sadri (Lyles School of Civil Engineering, Purdue University, 550 Stadium Mall Drive, West Lafayette, IN 47907, USA; Email: sadri.buet@gmail.com)

**Supplementary Figure S1 | Distribution of common users appeared in the most active day**

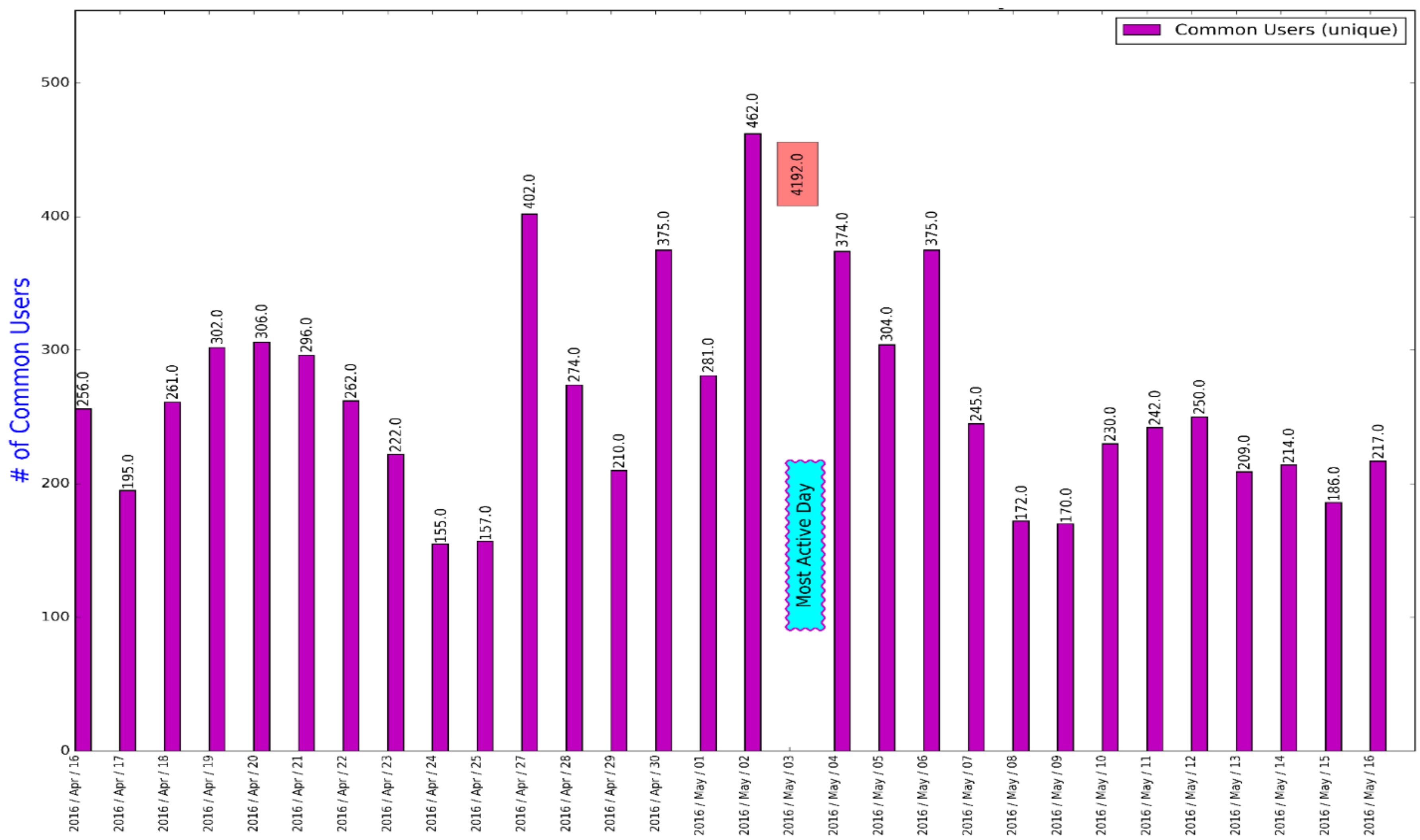

**Supplementary Figure S2 | Distribution of common links appeared in the most active day**

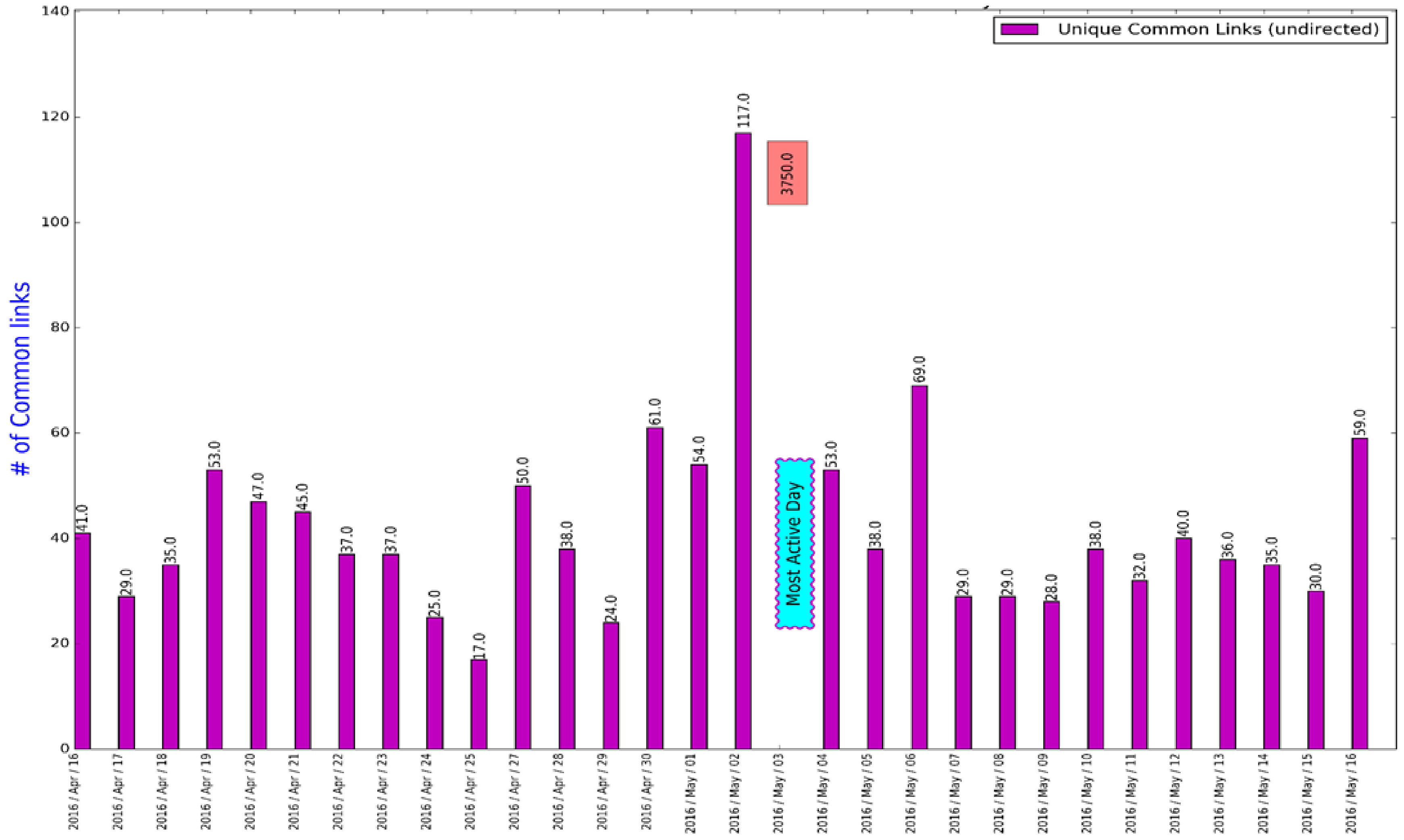

**Supplementary Figure S3 | Perplexity Analysis**

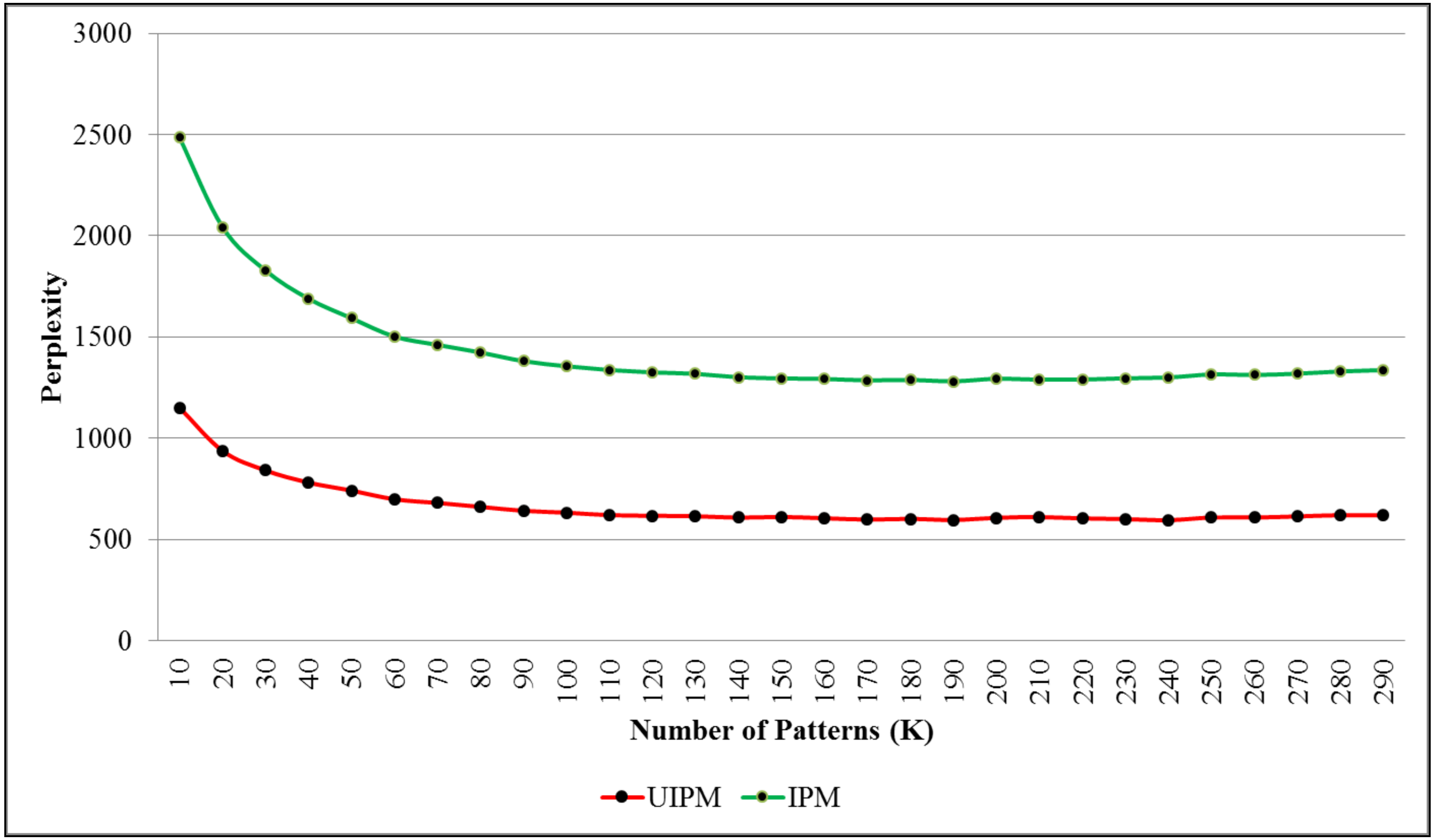

**Supplementary Figure S4 | Number of Users with Community Assignment**

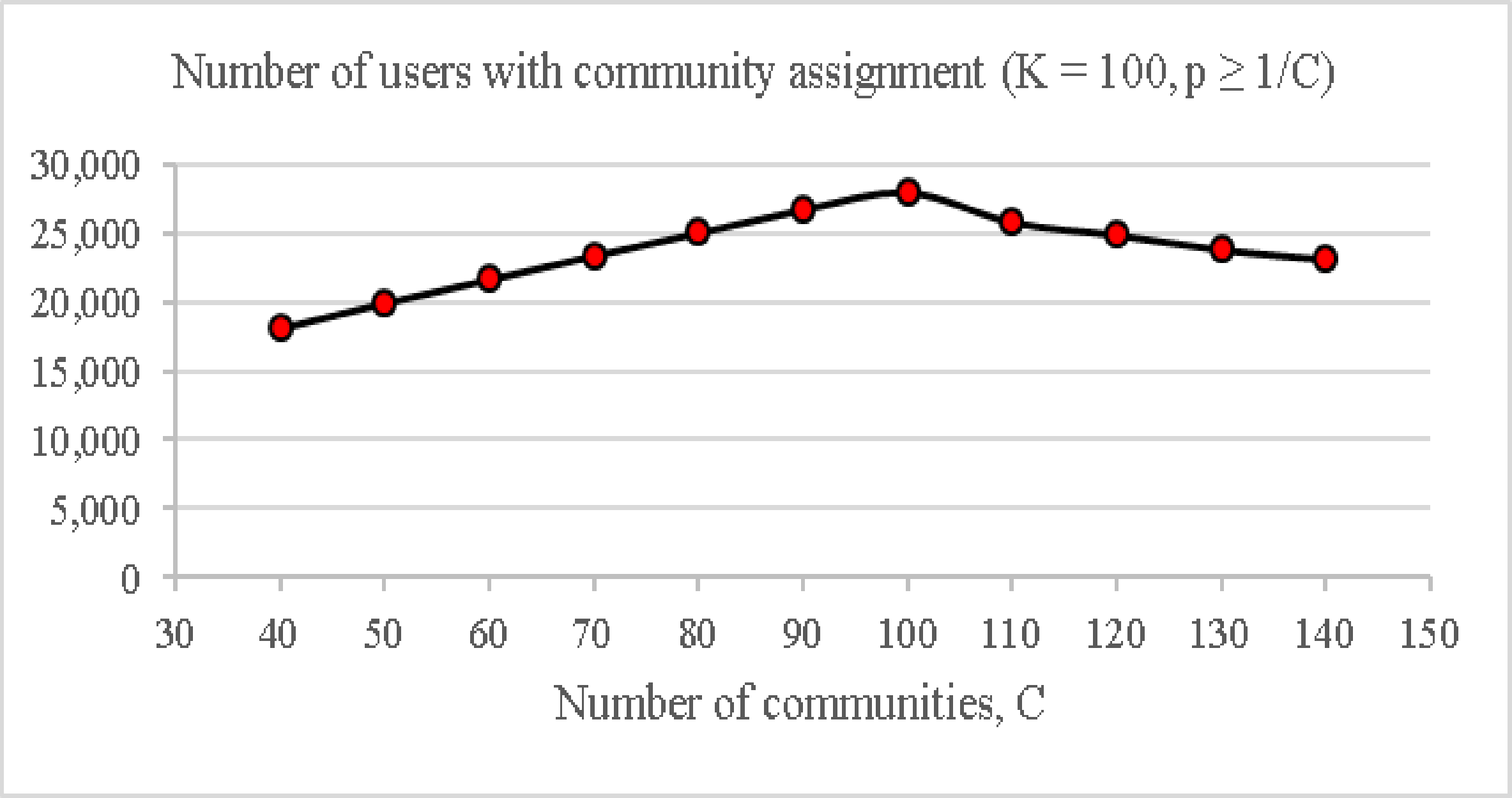

**Supplementary Figure S5 | Identifying user communities in the largest connected component using CIPM (a) original graph [all nodes color coded as white], (b) classified graph [color code: C-1 (red), C-21 (green), C-3 (blue) and C-2 (yellow)]**

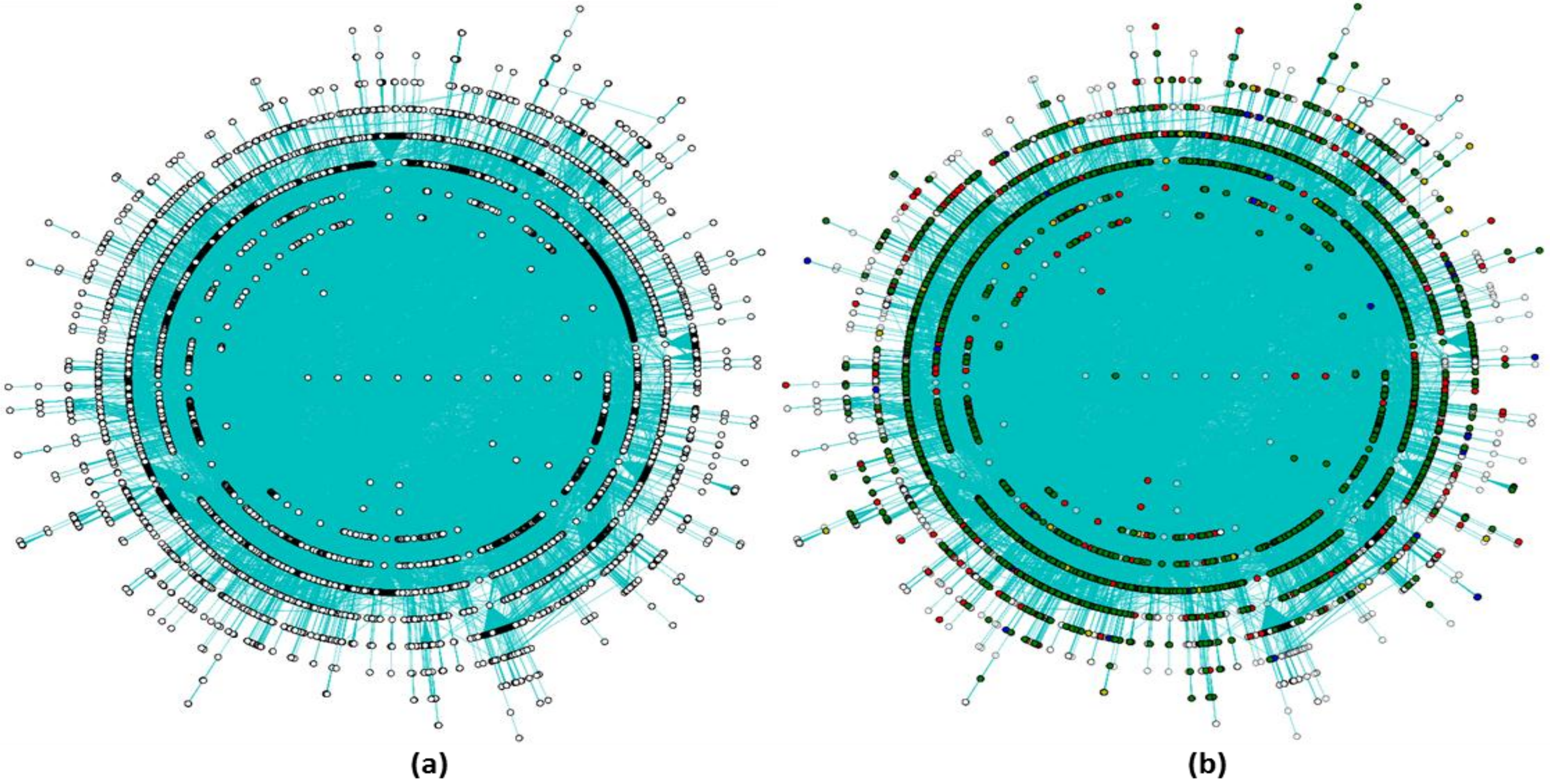

## Supplementary Table S1 | Interest pattern model (IPM) results

| *w* | *P(w\|Z* | *w* | *P(w\|Z* | *w* | *P(w\|Z* | *w* | *P(w\|Z* | *w* | *P(w\|Z* | *w* | *P(w\|Z* |
|---|---|---|---|---|---|---|---|---|---|---|---|
| ***Academic*** | | | | | | ***Alumni*** | | ***Graduation*** | | ***Tuition Fees*** | |
| **Pattern 1** | | **Pattern** | **0.01** | **Pattern** | | **Pattern 97** | | **Pattern 7** | | **Pattern 11** | |
| university | 0.14 | class | 0.07 | way | 0.06 | purdue | 0.20 | congrats | 0.08 | year | 0.07 |
| students | 0.08 | purdue | 0.06 | new | 0.06 | alumni | 0.03 | today | 0.07 | tuition | 0.04 |
| ive | 0.07 | fall | 0.03 | college | 0.05 | center | 0.02 | proud | 0.06 | get | 0.04 |
| three | 0.07 | coming | 0.03 | purdue | 0.04 | dance | 0.02 | engineering | 0.05 | people | 0.03 |
| far | 0.05 | boilerup | 0.03 | via | 0.04 | golf | 0.01 | congratulatio | 0.03 | hours | 0.03 |
| semester | 0.05 | new | 0.03 | education | 0.03 | fun | 0.01 | purduewedid | 0.03 | straight | 0.03 |
| gone | 0.04 | purdues | 0.03 | program | 0.02 | event | 0.01 | graduating | 0.02 | president | 0.03 |
| wonder | 0.04 | look | 0.02 | pay | 0.02 | got | 0.01 | graduation | 0.02 | didnt | 0.03 |
| failing | 0.04 | icymi | 0.02 | university | 0.01 | haas | 0.01 | son | 0.02 | daniels | 0.02 |
| lectures | 0.04 | 2020 | 0.02 | income | 0.01 | culture | 0.01 | graduates | 0.02 | 12 | 0.02 |
| ***Research & Innovation*** | | | | | | ***Donation*** | | ***Online Writing Lab*** | | ***Administration*** | |
| **Pattern** | | **Pattern** | **0.01** | **Pattern** | | **Pattern 23** | | **Pattern 2** | **0.01** | **Pattern 29** | |
| student | 0.08 | new | 0.06 | purdue | 0.09 | offer | 0.17 | owl | 0.03 | purdue | 0.11 |
| purdue | 0.06 | research | 0.06 | graduate | 0.06 | university | 0.10 | since | 0.03 | made | 0.08 |
| david | 0.04 | award | 0.04 | space | 0.05 | blessed | 0.10 | purdue | 0.02 | online | 0.05 |
| science | 0.03 | deal | 0.02 | grad | 0.04 | receive | 0.07 | check | 0.02 | employees | 0.03 |
| helps | 0.03 | cancer | 0.02 | mission | 0.03 | boilerup | 0.06 | use | 0.02 | software | 0.03 |
| hedelin | 0.03 | may | 0.02 | station | 0.03 | received | 0.06 | new | 0.02 | scheduling | 0.03 |
| data | 0.03 | 2016 | 0.01 | 2017 | 0.03 | say | 0.03 | writing | 0.02 | managing | 0.03 |
| success | 0.02 | program | 0.01 | nasa | 0.02 | honored | 0.02 | even | 0.02 | easier | 0.03 |
| ol | 0.02 | expand | 0.01 | join | 0.02 | thankful | 0.02 | career | 0.02 | guys | 0.02 |
| predict | 0.02 | read | 0.01 | tingle | 0.02 | god | 0.01 | follow | 0.02 | days | 0.01 |
| ***Purdue Day of Giving*** | | | | | | ***Indiana Primary Election 2016*** | | | | | |
| **Pattern 5** | | **Pattern** | | **Pattern** | | **Pattern 78** | | **Pattern 82** | | **Pattern 85** | |
| dayofgivi | 0.16 | today | 0.14 | future | 0.08 | sanders | 0.13 | purdue | 0.30 | indiana | 0.09 |
| us | 0.06 | dayofgivi | 0.08 | purdue | 0.07 | bernie | 0.11 | w | 0.04 | indy | 0.07 |
| help | 0.06 | better | 0.05 | igave | 0.05 | hall | 0.04 | 1st | 0.02 | notredame | 0.06 |
| support | 0.06 | making | 0.04 | grant | 0.04 | berniesande | 0.04 | place | 0.02 | inprimary | 0.06 |
| make | 0.03 | join | 0.04 | opportunit | 0.04 | town | 0.04 | trump | 0.01 | feelthebern | 0.06 |
| donate | 0.03 | bigger | 0.03 | student | 0.04 | rally | 0.03 | support | 0.01 | iu | 0.06 |
| give | 0.02 | supportin | 0.02 | could | 0.04 | live | 0.03 | agree | 0.01 | indianastat | 0.05 |
| pubandorc | 0.02 | graduatin | 0.02 | shape | 0.03 | line | 0.02 | chalk | 0.01 | isu | 0.05 |
| gift | 0.02 | pmo | 0.02 | dayofgivi | 0.02 | political | 0.01 | days | 0.01 | vicaucus | 0.05 |
| hour | 0.02 | upon | 0.02 | dayofgivi | 0.02 | revolution | 0.01 | idc | 0.01 | virginislan | 0.04 |

## Supplementary Table S2 | Interest pattern model (IPM) results (continued)

| *w* | *P(w\|Z)* | *w* | *P(w\|Z)* | *w* | *P(w\|Z)* | *w* | *P(w\|Z)* | *w* | *P(w\|Z)* | *w* | *P(w\|Z)* |
|---|---|---|---|---|---|---|---|---|---|---|---|
| ***Celebrities/Players*** | | | | | | | | | | | |
| ***Danny Anthrop*** | | ***Caleb Swanigan*** | | ***Champion Villanova*** | | ***Coach Gene Keady*** | | ***Brown Anthony*** | | ***Spike Albrecht*** | |
| **Pattern 12** | | **Pattern 15** | | **Pattern 17** | | **Pattern 33** | | **Pattern 64** | | **Pattern 36** | |
| state | 0.04 | purdues | 0.13 | national | 0.06 | coach | 0.09 | anthony | 0.10 | spike | 0.12 |
| illinois | 0.03 | nba | 0.07 | host | 0.06 | keady | 0.08 | brown | 0.10 | albrecht | 0.12 |
| anthrop | 0.03 | swanigan | 0.06 | villanova | 0.04 | gene | 0.07 | cb | 0.09 | transfer | 0.11 |
| danny | 0.03 | caleb | 0.05 | breaking | 0.04 | former | 0.07 | cowboys | 0.06 | season | 0.07 |
| list | 0.02 | combine | 0.04 | defending | 0.04 | basketball | 0.06 | dallas | 0.04 | michigan | 0.07 |
| iowa | 0.02 | draft | 0.03 | 14.0000 | 0.04 | rally | 0.05 | pick | 0.04 | play | 0.06 |
| butler | 0.02 | wants | 0.02 | nov | 0.03 | took | 0.04 | four | 0.03 | next | 0.05 |
| among | 0.02 | invited | 0.02 | gavittgames | 0.03 | longtime | 0.04 | round | 0.03 | final | 0.04 |
| free | 0.02 | players | 0.02 | 6.0000 | 0.02 | carmel | 0.04 | 6th | 0.02 | pg | 0.03 |
| colts | 0.02 | hammons | 0.02 | nfldraft | 0.02 | stage | 0.04 | draft | 0.02 | eligible | 0.02 |
| ***Games*** | | | | | | | | | | | |
| ***Ohio Gobucks*** | | ***Minnesota Gophers*** | | ***Game with Other Teams*** | | ***Big Ten Conference/NCAA*** | | | | | |
| **Pattern 28** | | **Pattern 35** | | **Pattern 53** | | **Pattern 81** | | **Pattern 14** | | **Pattern 77** | |
| win | 0.08 | 3 | 0.07 | vs | 0.10 | purdue | 0.17 | boilermakers | 0.15 | purdue | 0.21 |
| gobucks | 0.04 | final | 0.04 | game | 0.07 | 2 | 0.08 | university | 0.09 | school | 0.08 |
| b1g | 0.03 | series | 0.03 | maryland | 0.04 | vs | 0.04 | purdue | 0.05 | high | 0.02 |
| ohio | 0.03 | 1 | 0.03 | football | 0.04 | saturday | 0.04 | bigten | 0.05 | ready | 0.02 |
| buckeyes | 0.03 | innings | 0.02 | 2016 | 0.03 | field | 0.04 | ncaa | 0.02 | head | 0.02 |
| sweep | 0.02 | minnesota | 0.02 | terps | 0.03 | tbt | 0.02 | black | 0.02 | ncaagolf | 0.02 |
| finish | 0.02 | 12 | 0.02 | 1 | 0.03 | pregame | 0.02 | pete | 0.01 | practice | 0.01 |
| recap | 0.02 | gophers | 0.02 | softball | 0.02 | wvu | 0.02 | large | 0.01 | polytechnic | 0.01 |
| baseball | 0.02 | top | 0.02 | announced | 0.02 | sept | 0.02 | jersey | 0.01 | speak | 0.01 |
| state | 0.02 | runs | 0.02 | feartheturtle | 0.02 | mountaineer | 0.01 | logo | 0.01 | stem | 0.01 |
| ***Purdue Pharma*** | | | | | | | | | | | |
| **Pattern 19** | | **Pattern 24** | | **Pattern 34** | | **Pattern 66** | | **Pattern 86** | | **Pattern 90** | |
| epidemic | 0.07 | purdue | 0.13 | purdue | 0.26 | oxycontin | 0.07 | times | 0.08 | oxycontin | 0.08 |
| helped | 0.07 | oxycontins | 0.04 | miss | 0.03 | pharma | 0.05 | oxycontin | 0.08 | records | 0.06 |
| opioid | 0.07 | 12hour | 0.02 | says | 0.02 | next | 0.04 | pharma | 0.06 | maker | 0.06 |
| pharma | 0.06 | problem | 0.02 | houston | 0.01 | judge | 0.03 | la | 0.05 | keep | 0.05 |
| via | 0.04 | hell | 0.02 | doctors | 0.01 | secret | 0.03 | got | 0.04 | painkiller | 0.04 |
| drug | 0.04 | want | 0.02 | misled | 0.01 | week | 0.03 | says | 0.04 | block | 0.04 |
| companies | 0.04 | marketing | 0.01 | yesterday | 0.01 | records | 0.02 | heres | 0.04 | secret | 0.03 |
| cause | 0.03 | former | 0.01 | friday | 0.01 | documents | 0.02 | report | 0.04 | lawsuit | 0.03 |
| profits | 0.03 | claim | 0.01 | pharmaceuticals | 0.01 | unseal | 0.02 | investigation | 0.03 | battle | 0.03 |
| evidence | 0.03 | created | 0.01 | cold | 0.01 | thanks | 0.01 | wrong | 0.02 | legal | 0.03 |

## Supplementary Table S2 | User Interest Pattern Model (UIPM) results

***Indiana Primary Election 2016***

| Pattern 1 | 0.0136 | Pattern 56 | 0.0142 | Pattern 21 | 0.0117 |
|---|---|---|---|---|---|
| ***w*** | ***P(w\|Z)*** | ***w*** | ***P(w\|Z)*** | ***w*** | ***P(w\|Z)*** |
| sanders | 0.1229 | indiana | 0.0858 | purdue | 0.0768 |
| bernie | 0.0918 | feelthebern | 0.0579 | berniesanders | 0.0424 |
| live | 0.0283 | notredame | 0.0557 | line | 0.0223 |
| rally | 0.0237 | inprimary | 0.0550 | hall | 0.0167 |
| town | 0.0237 | iu | 0.0487 | town | 0.0125 |
| hall | 0.0190 | indy | 0.0484 | rally | 0.0122 |
| win | 0.0151 | indianastate | 0.0474 | today | 0.0107 |
| crowd | 0.0143 | purdue | 0.0435 | purdues | 0.0095 |
| become | 0.0121 | vicaucus | 0.0425 | night | 0.0080 |
| revolution | 0.0117 | virginislands | 0.0410 | last | 0.0079 |
| ***u*** | ***P(u\|Z)*** | ***u*** | ***P(u\|Z)*** | ***u*** | ***P(u\|Z)*** |
| 41621505 | 0.0366 | 16664309 | 0.1078 | 11775782 | 0.0499 |
| 16440677 | 0.0275 | 721428816971894000 | 0.0185 | 158216203 | 0.0486 |
| 2228806220 | 0.0189 | 123269504 | 0.0087 | 463204470 | 0.0287 |
| 354330741 | 0.0152 | 2416940856 | 0.0078 | 2350966291 | 0.0180 |
| 88245817 | 0.0122 | 2332808262 | 0.0065 | 199636296 | 0.0123 |
| 802584787 | 0.0100 | 713948924004831000 | 0.0063 | 376476523 | 0.0116 |
| 1952653698 | 0.0056 | 4871145023 | 0.0062 | 705454296154796000 | 0.0106 |
| 242137063 | 0.0053 | 4582019741 | 0.0058 | 3056052077 | 0.0104 |
| 463204470 | 0.0048 | 1860727694 | 0.0051 | 53983 | 0.0075 |
| 50793235 | 0.0036 | 24190388 | 0.0044 | 2333439294 | 0.0067 |

***Purdue Day of Giving***

| Pattern 19 | 0.0231 | Pattern 20 | 0.0092 | Pattern 25 | 0.0084 |
|---|---|---|---|---|---|
| ***w*** | ***P(w\|Z)*** | ***w*** | ***P(w\|Z)*** | ***w*** | ***P(w\|Z)*** |
| dayofgiving | 0.1358 | purdue | 0.1649 | purdue | 0.1701 |
| support | 0.0322 | day | 0.1101 | pubandorch | 0.0442 |
| help | 0.0304 | giving | 0.0512 | purduedayofgiving | 0.0296 |
| today | 0.0253 | pmo | 0.0184 | boiler | 0.0268 |
| igave | 0.0219 | today | 0.0160 | black | 0.0243 |
| us | 0.0215 | msu | 0.0117 | band | 0.0217 |
| better | 0.0195 | million | 0.0113 | hail | 0.0189 |
| grant | 0.0195 | record | 0.0104 | gold | 0.0152 |
| future | 0.0175 | ever | 0.0101 | big | 0.0112 |
| make | 0.0175 | lets | 0.0095 | marching | 0.0102 |
| ***u*** | ***P(u\|Z)*** | ***u*** | ***P(u\|Z)*** | ***u*** | ***P(u\|Z)*** |
| 28573278 | 0.0293 | 772180069 | 0.0746 | 215702601 | 0.1082 |
| 16134525 | 0.0142 | 3525080836 | 0.0083 | 381639855 | 0.0233 |
| 605564976 | 0.0126 | 110691113 | 0.0048 | 71036725 | 0.0123 |
| 45052134 | 0.0122 | 11775782 | 0.0045 | 14421621 | 0.0101 |
| 772180069 | 0.0108 | 95450798 | 0.0039 | 66167575 | 0.0062 |
| 348440739 | 0.0106 | 348440739 | 0.0038 | 724630593800179000 | 0.0059 |
| 829194631 | 0.0095 | 902520864 | 0.0033 | 1520322312 | 0.0049 |
| 44409797 | 0.0084 | 190252722 | 0.0033 | 4864158051 | 0.0048 |
| 3420171568 | 0.0082 | 369128406 | 0.0029 | 2533104662 | 0.0045 |
| 3215171243 | 0.0080 | 312050513 | 0.0027 | 3313270876 | 0.0043 |

# Supplementary Table S3 | Community Interest Pattern Model (CIPM) results

| *w* | *P(w\|Z)* | *w* | *P(w\|Z)* | *w* | *P(w\|Z)* |
|---|---|---|---|---|---|
| **Pattern 1** | 0.146 | **Pattern 5** | 0.006 | **Pattern 23** | 0.009 |
| purdue | 0.175 | way | 0.031 | visit | 0.075 |
| university | 0.023 | new | 0.030 | weekend | 0.031 |
| game | 0.008 | could | 0.022 | transfer | 0.030 |
| boilerup | 0.008 | college | 0.018 | told | 0.026 |
| today | 0.007 | based | 0.017 | tech | 0.026 |
| coach | 0.006 | long | 0.016 | stephens | 0.020 |
| spring | 0.006 | forget | 0.014 | plans | 0.019 |
| first | 0.006 | program | 0.014 | kendall | 0.019 |
| great | 0.006 | never | 0.014 | take | 0.018 |
| team | 0.006 | pay | 0.013 | texas | 0.018 |
| **Pattern 62** | 0.008 | **Pattern 91** | 0.007 | **Pattern 99** | 0.014 |
| like | 0.030 | purdue | 0.128 | michigan | 0.048 |
| recruiting | 0.029 | vs | 0.024 | iowa | 0.044 |
| iubb | 0.028 | softball | 0.023 | offers | 0.037 |
| taking | 0.026 | 2018 | 0.022 | st | 0.030 |
| iu | 0.022 | play | 0.018 | state | 0.029 |
| matt | 0.018 | thank | 0.016 | interest | 0.025 |
| lot | 0.017 | tyler | 0.015 | wisconsin | 0.021 |
| story | 0.016 | head | 0.012 | week | 0.019 |
| robert | 0.016 | jeff | 0.011 | msu | 0.019 |
| hes | 0.015 | g | 0.011 | illinois | 0.017 |

| **Community 2 (LCC)** (p>=0.025) | **Topics of maximum interest** | | **Community 3 (LCC)** (p>=0.025) | **Topics of maximum interest** | |
|---|---|---|---|---|---|
| **User Id** | ***Pattern*** | ***p*** | **User Id** | ***Pattern*** | ***p*** |
| 3234688946 | 1 | 0.200 | 615833064 | 1 | 0.461 |
| 3924021759 | 1 | 0.396 | 2710709881 | 1 | 0.321 |
| 537760351 | 1 | 0.615 | 1069622622 | 1 | 0.299 |
| 3404906992 | 1 | 0.164 | 90658098 | 1 | 0.206 |
| 1049183928 | 1 | 0.246 | 1180961940 | 1 | 0.288 |
| 62787754 | 1 | 0.271 | 23646710 | 23 | 0.237 |
| 622150102 | 1 | 0.280 | 3281659363 | 23 | 0.204 |
| 1241975864 | 1 | 0.307 | 168906812 | 62 | 0.360 |
| 913748036 | 1 | 0.169 | 304160410 | 62 | 0.214 |
| 3015956952 | 1 | 0.277 | 70952154 | 91 | 0.309 |
| 714816804858630144 | 5 | 0.462 | 21438334 | 99 | 0.476 |